\documentclass[%
 reprint,
 floatfix,
superscriptaddress,
 amsmath,amssymb,
 aps
]{revtex4-2}
\usepackage{xcolor}
\usepackage{graphicx}
\usepackage{dcolumn}
\usepackage{bm}
\usepackage{subfig}
\usepackage{soul}
\usepackage{comment}
\begin{document}

\preprint{APS/123-QED}

\title{Collision-induced C$_{60}$ rovibrational relaxation probed by state-resolved nonlinear spectroscopy}% Force line breaks with \\

\author{Lee R. Liu}
\email{lee.richard.liu@gmail.com}
\affiliation{JILA, National Institute of Standards and Technology and University of Colorado, Boulder, Boulder, CO 80309}
\affiliation{Department of Physics, University of Colorado, Boulder, CO 80309}

\author{P. Bryan Changala}
\altaffiliation[Present address: ]{Center for Astrophysics, Harvard \& Smithsonian, Cambridge, MA 02138}
\affiliation{JILA, National Institute of Standards and Technology and University of Colorado, Boulder, Boulder, CO 80309}

\author{Marissa L. Weichman}
\altaffiliation[Present address: ]{Department of Chemistry, Princeton University, Princeton, NJ 08544}
\affiliation{JILA, National Institute of Standards and Technology and University of Colorado, Boulder, Boulder, CO 80309}

\author{Qizhong Liang}
\affiliation{JILA, National Institute of Standards and Technology and University of Colorado, Boulder, Boulder, CO 80309}
\affiliation{Department of Physics, University of Colorado, Boulder, CO 80309}

\author{Jutta Toscano}
\altaffiliation[Present address: ]{Department of Chemistry, University of Basel, 4056 Basel, Switzerland}
\affiliation{JILA, National Institute of Standards and Technology and University of Colorado, Boulder, Boulder, CO 80309}

\author{Jacek K\l{}os}
\affiliation{Department of Physics, Temple University, Philadelphia, PA 19122}

\author{Svetlana Kotochigova}
\affiliation{Department of Physics, Temple University, Philadelphia, PA 19122}

\author{David J. Nesbitt}
\affiliation{JILA, National Institute of Standards and Technology and University of Colorado, Boulder, Boulder, CO 80309}
\affiliation{Department of Physics, University of Colorado, Boulder, CO 80309}
\affiliation{Department of Chemistry, University of Colorado, Boulder, CO 80309}

\author{Jun Ye}
\email{ye@jila.colorado.edu}
\affiliation{JILA, National Institute of Standards and Technology and University of Colorado, Boulder, Boulder, CO 80309}
\affiliation{Department of Physics, University of Colorado, Boulder, CO 80309}

\date{\today}% It is always \today, today,

\begin{abstract}
Quantum state-resolved spectroscopy was recently achieved for C$_{60}$ molecules when cooled by buffer gas collisions and probed with a mid-infrared frequency comb. This rovibrational quantum state resolution for the largest molecule on record is facilitated by the remarkable symmetry and rigidity of C$_{60}$, which also present new opportunities and challenges to explore energy transfer between quantum states in this many-atom system. Here we combine state-specific optical pumping, buffer gas collisions, and ultrasensitive intracavity nonlinear spectroscopy to initiate and probe the rotation-vibration energy transfer and relaxation. This approach provides the first detailed characterization of C$_{60}$ collisional energy transfer for a variety of collision partners, and determines the rotational and vibrational inelastic collision cross sections. These results compare well with our theoretical modelling of the collisions, and establish a route towards quantum state control of a new class of unprecedentedly large molecules.
\end{abstract}

\maketitle

%\tableofcontents

\section{\label{sec:level1}Introduction}

Understanding the excitation and relaxation pathways of complex quantum mechanical systems is a primary focus of chemical and many-body physics. Precision frequency-domain spectroscopy provides a unique probe of these dynamics, elucidating structure and interactions at the level of individual molecular quantum states. High-resolution spectroscopic investigations, however, have traditionally been limited to relatively small molecules containing fewer than a dozen atoms, due to the challenges of creating cold, controlled gas-phase samples and the intrinsic spectral congestion of larger systems. 

Recently, these obstacles have been overcome for large molecules, including buckminsterfullerene C$_{60}$, by a combination of two key experimental techniques: i) cryogenic buffer-gas cooling, which generates a sample of cold, gas-phase molecules, and ii) cavity-enhanced infrared frequency comb spectroscopy, which probes the molecular rovibrational spectrum with high sensitivity and spectral resolution~\cite{Spaun2016,Changala2019}. Although the initial study of C$_{60}$ has revealed structural information such as rotational constants, vibrational energy spacings, and Coriolis interaction strength, the small optical power available per frequency comb component limits this approach as a fundamentally passive technique.  

Further insight into the internal dynamics of large molecular systems requires the understanding and control of inherent decoherence processes.  In this work, state-resolved optical pumping with a single-frequency laser reveals the existence of such processes in collisional interactions of C$_{60}$ with atoms and diatomic molecules. Our experiments and models provide guidance for mitigating these sources of decoherence, as well as for enhancing efficiency of collisional cooling. Ultimately, we will need to gain control over the internal and external degrees of freedom of individual C$_{60}$ molecules in order to probe novel quantum dynamics with sixty interconnected C atoms.

\section{Experimental Details}
In order to manipulate rovibrational state populations and determine transition lineshapes with greatly enhanced sensitivity and precision, we demonstrate a new scheme that employs a continuous wave (cw) laser of enhanced optical power at a specific probe frequency. Coupling a quantum cascade laser (QCL) to a high finesse optical cavity increases the probe intensity on cold C$_{60}$ by $10^4$-fold, reaching well into the saturated absorption regime where an appreciable internal state population is driven out of thermal equilibrium. Further, the concomitant 100-fold gain in sensitivity reveals detailed lineshape profiles that furnish a wealth of information on relaxation and diffusion~\cite{Rautian1967,Galatry1961,Varghese1984a,Rohart2008,Nienhuis1978,Williams1973,Weber1981,Ciuryo1998,Schiffman1994}. We measure saturated absorption lineshape profiles and compare them to a detailed rate-equation model, thereby mapping out the propensity of C$_{60}$ for inter-converting energy from collisions with the surrounding bath into translation and internal vibrations and rotations.

\begin{figure*}
	\includegraphics[width=2\columnwidth,scale=1]{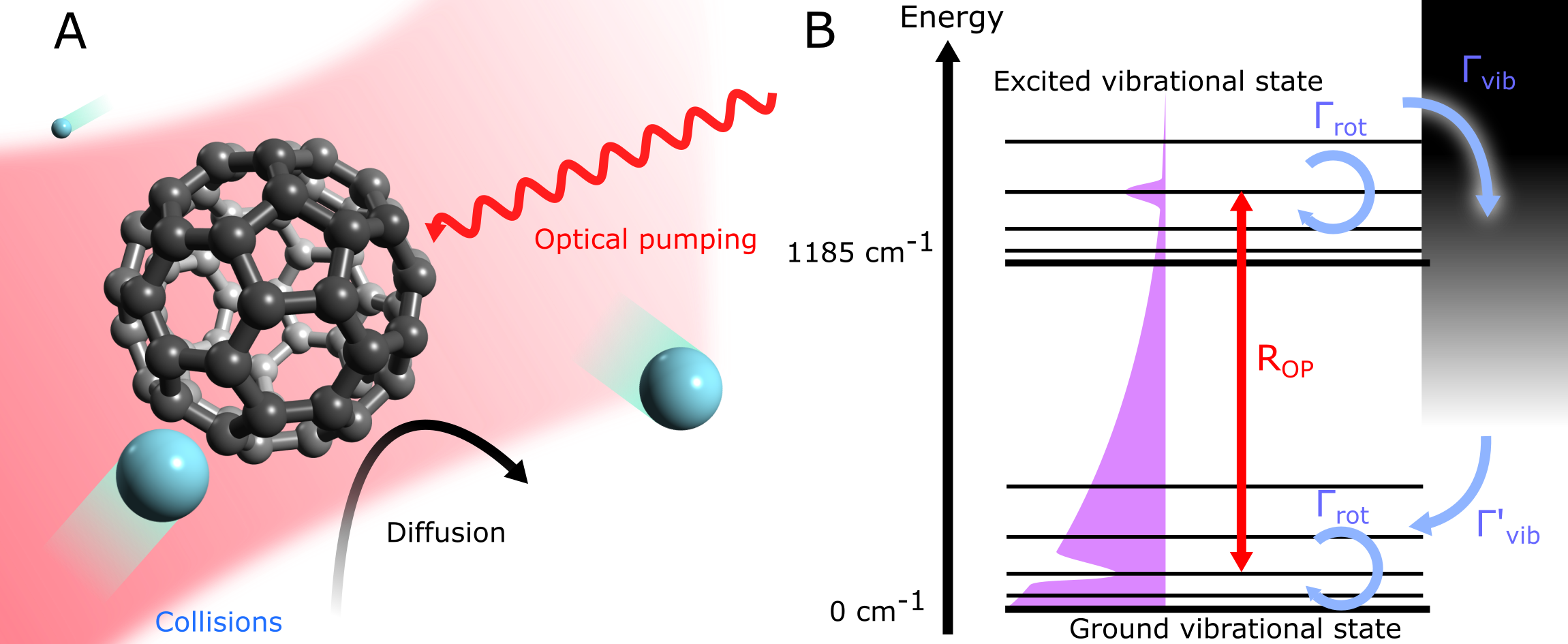}
	\caption{Pumping and collisional relaxation processes in buffer gas (BG)-cooled C$_{60}$. A) Interactions of C$_{60}$ with its environment. C$_{60}$ diffuses into the optical cavity mode, where it is pumped by the cw intracavity field before diffusing out again. The experiment monitors the frequency-dependent nonlinear absorption of the intracavity field by C$_{60}$. Collisions with the BG control diffusion and thermalize the internal rovibrational energy of C$_{60}$. B) Energy-level diagram of C$_{60}$ undergoing interactions with the environment. Optical pumping at a rate R$_{OP}$ drives the 1185~cm$^{-1}$($8.4~\mu$m) $T_{1u}(3)$ vibrational band of C$_{60}$. Collisions with the BG redistribute rotational population at a rate $\Gamma_{rot}$, assumed to be identical in $v''=0$ and $v'=1$. Vibrational relaxation is modeled in two steps: loss of vibrational population from $v'=1$ into the dense reservoir of dark vibrational states at a rate $\Gamma_{vib}$, and leakage from the dense reservoir into $v''=0$ at a rate $\Gamma_{vib}'$. Filled purple curve depicts the steady-state population distribution.}
	\label{fig:microscopic_picture}
\end{figure*} 

Fig.~\ref{fig:microscopic_picture}A depicts the physical system under consideration. A buffer gas (BG) cooling cell is nested inside an enhancement cavity with finesse of 12,000. The cw-QCL is coupled into this cavity to enhance both intensity for nonlinear spectroscopy and detection sensitivity. Further experimental details on laser stabilization and spectroscopy scanning are provided in the Methods section. Inside the cold cell, C$_{60}$ collides with the surrounding BG and, as it diffuses through the cavity mode, is simultaneously pumped and probed by the intracavity field with photon scattering rate $R_{OP}$. We neglect C$_{60}-$C$_{60}$ collisions because they are expected to be six orders of magnitude less frequent than C$_{60}-$BG collisions (Appendix ~\ref{sec:c60_c60_collisions}). 

\section{Optical pumping and collision dynamics}
The effects on the internal state populations of the ensemble of C$_{60}$ molecules are illustrated in Fig.~\ref{fig:microscopic_picture}B. Continuous optical pumping drives the system out of thermal equilibrium, towards equal populations of the connected states and therefore a reduced absorption cross section.  Simultaneously, thermal equilibrium is restored by collisions in two ways. First, inelastic collisions induce rotational and vibrational transitions at rates $\Gamma_{rot}$ and $\Gamma_{vib}$. Second, collisions control the rate of diffusion, which replaces optically pumped molecules in the cavity mode with thermal molecules from outside the cavity mode. In steady state, the saturation of the absorption cross section therefore provides a direct comparison between $R_{OP}$ and the various rates of collisional thermalization~\cite{Demtroder2013}.

\begin{figure*}
	\includegraphics[width=1.5\columnwidth,scale=1]{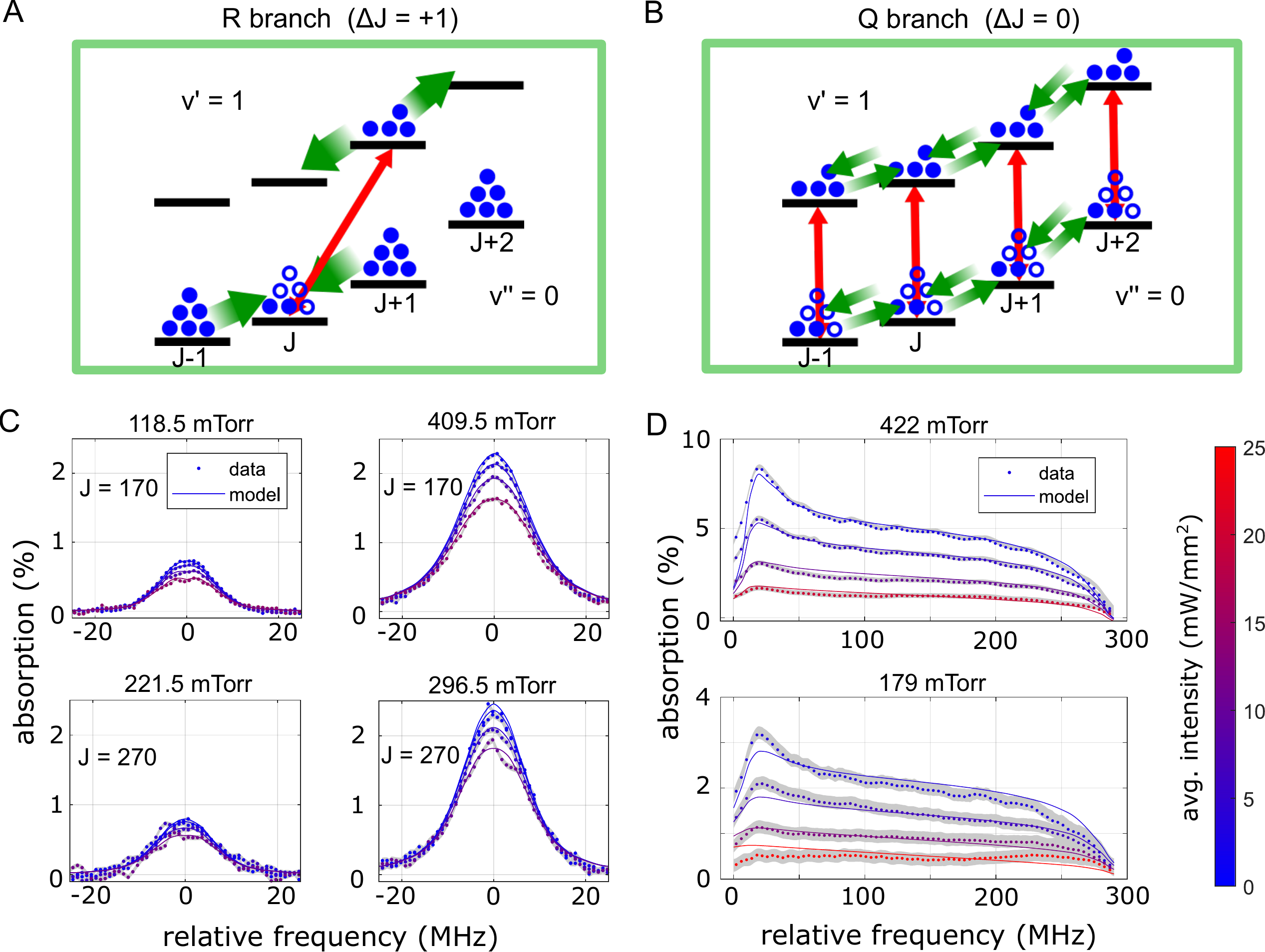}
	\caption{Effect of single-state versus multi-state pumping on rotational relaxation in C$_{60}$. A) State-selective pumping in the R-branch. Only a single transition can be resonantly pumped at a time; therefore, rotational relaxation from C$_{60}$-BG collisions efficiently redistributes pumped rotational populations (green arrows). B) Multi-state pumping in the Q-branch, with many transitions (up to $\sim 75$ at the band-head) being pumped simultaneously. C) Rate equation model fits to C$_{60}-$Ar R-branch saturated absorption data. Sampling of R-branch absorption data at $J=170$ (top) and $J=270$ (bottom), and corresponding rate equation fits (full C$_{60}-$Ar R-branch data in Appendix~\ref{sec:C60-Ar_R_full}). Good agreement is obtained by modeling state-to-state rotational relaxation with an exponential gap law (see text). Etalons have been fitted out and subtracted from the data. Gray bands indicate the 1-sigma uncertainty in the data. D) Rate equation model fits to C$_{60}-$Ar Q-branch saturated absorption data. Q branch absorption data and rate equation model fit (described in text) taken at high (upper panel) and low (lower panel) Ar pressure (specified above each panel) at four pumping intensities. Only every fourth data point is plotted for clarity. Gray bands indicate standard uncertainty in the data dominated by residual etalons in the optical setup. The color of traces in panels C) and D) correspond to different pumping intensities, indicated by the color bar.} 
	\label{fig:saturation_Q_R}
\end{figure*}

To distinguish the effect of $\Gamma_{rot}$ and $\Gamma_{vib}$, we probe and compare single- and multi-state pumping of C$_{60}$ enabled by the R and Q branch transitions respectively, as shown in Fig.~\ref{fig:saturation_Q_R}A and B. In the R branch, $\Delta J = +1$  transitions are widely spaced relative to their line width in a regular progression; therefore, only one transition is resonantly pumped at a time. Rotational relaxation from C$_{60}-$BG collisions redistributes pumped rotational populations into empty neighbouring states as well as replenishes the lower state population, so that the degree of saturation measured under R-branch pumping is highly sensitive to both vibrational and rotational relaxation rates. On the other hand, the spectrally overlapping $\Delta J =0 $ transitions of the Q branch permit multiple transitions to be pumped simultaneously. This renders the saturation intensity insensitive to the rotational redistribution rate, $\Gamma_{rot}$. Thus, the combination of Q and R branch measurements allows us to distinguish between relaxation due to rotationally and vibrationally inelastic collisions.

The saturation effect is parameterized with $s_0=i/i_{sat}$, where $i$ is the mean intracavity intensity and $i_{sat}$ is the saturation intensity. We determine $s_0$ by measuring absorption profiles as a function of $i=2 - 25$~mW/mm$^2$. At a typical background pressure of 300~mTorr of argon gas, this corresponds to $s_0 = 0.03 - 0.3$ in the R branch. Sample nonlinear spectra of the R and Q branches of C$_{60}$ in Ar buffer gas are shown in  Figure ~\ref{fig:saturation_Q_R} C and D, respectively. As expected, the R branch shows a well-resolved Voigt line shape for each $\Delta J=+1$ transition, while the Q branch appears as an unresolved contour due to spectrally overlapping $\Delta J=0$ transitions. As discussed further below, the Voigt profiles are dominated by homogeneous (pressure-broadening) and inhomogeneous (Doppler broadening) widths of around $10~$MHz each, while transit time broadening is expected to only be a few kHz. The band head in the Q branch contour arises from centrifugal distortion effects causing several $\Delta J=0$ transitions ($J''\sim275 - 352$) to lie within one homogeneous linewidth of each other~\cite{Changala2019}.

Notably, the Q branch is much more strongly saturated, as expected from the higher effective absorption cross section and suppression of rotational population redistribution.
%peak R-branch photon scattering rates $R_{OP} = 3\times10^4-3\times10^5$~s$^{-1}$, or the 
Since the diffusion rate (parameterized by the ensemble-averaged total cross section $\overline{\sigma}_{tot}$; see Appendix~\ref{sec:diff_const}) increases at lower pressures, while inelastic collision rates (parameterized by rotationally and vibrationally inelastic cross sections $\sigma_{rot}$ and $\sigma_{vib}$) correspondingly decrease, measurement of the saturated Q and R branch absorption profiles over a range of BG pressures permits a complete map of energy flow from a collision to the various internal modes of C$_{60}$.

%and the transverse diffusion rate out of the beam volume $A_{diff} = 2\frac{\langle v_{rel} \rangle}{n \sigma_{tot} w_0^2}$~\cite{SI}

%Here, we have defined the buffer gas number density $n = \frac{P}{kT}$,  thermally averaged C$_{60}$-BG relative speed $\langle v_{rel} \rangle=\sqrt{\frac{8 k T}{\pi \mu}}$, and the characteristic length scale of the optical pumping region (here, the calculated cavity mode beam waist) $w_0=1.1~$mm. 
%$D=\frac{1}{2} \frac{\langle v_{rel} \rangle}{n \sigma_{tot}} $ is the diffusion constant .  

\section{Rate Equation Model}
The foregoing discussion provides a qualitative picture of how saturated absorption measurements at suitable pressures and pumping frequencies yield $\overline{\sigma}_{tot}$, $\sigma_{rot}$, and $\sigma_{vib}$. To obtain quantitative values we construct a rate equation model that simultaneously accounts for the evolution of states with total angular momentum $J$ ranging from $0~-~ 470$ in the ground vibrational, and triply degenerate first excited vibrational states (Appendix~\ref{sec:rate_eqn_model}), enabling a least-squares fit of all frequency-dependent saturation profiles simultaneously (solid lines in Fig.~\ref{fig:saturation_Q_R}C, D). A key assumption is that the translational, rotational, and vibrational degrees of freedom of C$_{60}$ are completely thermalized to the BG and cell wall in the absence of optical pumping. This is supported by the fitted Doppler widths of our R branch line shapes, rotational Boltzmann distribution observed in our previous work~\cite{Changala2019}, and the good absorption contrast which indicates that C$_{60}$ is principally in the ground vibrational state.

We first consider isolating the rotational population dynamics by applying an exponential-gap state-to-state rotational inelastic cross section fitting law~\cite{Polanyi1972,Bernstein1974,Lawley1877} to a full set of $J$-resolved R branch data over $J=110 - 270$ (Appendix~\ref{sec:C60-Ar_R_full}). We find an excellent fit across the entire range of measured $J$. The fitting law takes the form \begin{equation}
    \sigma_{rot}(J_f,J_i) =\alpha g_{f} e^{- \Delta E_{fi}/kT}
\end{equation} 
where $J_i,J_f$ are the initial and final $J$-states in the same vibrational manifold, $g_f$ is the angular momentum degeneracy of the final state calculated for the icosahedral spherical top~\cite{Bunker1980}, $\Delta E_{fi}=|E_f-E_i|$ is the absolute energy difference of the initial and final states, and $\alpha$ is a free scale parameter proportional to the thermally averaged integral cross section $\langle \sigma_{rot} \rangle$: 
\begin{align}\label{eq:sigma_rot}
    \langle \sigma_{rot} \rangle &\equiv \sum_{i,f\neq i} P(J_i) \sigma_{rot}(J_f,J_i)\\
    &= \alpha  \sum_{i,f\neq i} P(J_i) g_f e^{-\Delta E_{fi}/kT}\end{align} 
where 
\begin{equation}
P(J_i)=\frac{g_i}{Z_{rot}}~\textrm{exp}(-B J_i(J_i+1)/kT)    
\end{equation}
is the normalized rotational Boltzmann weight of the initial state with rotational partition function $Z_{rot}$. We assume that $\sigma_{rot}(J_f,J_i)$ and therefore $\langle \sigma_{rot} \rangle$ are identical in the ground and excited vibrational states.

%, which approaches $(2 J_f+1)^2/60$ for large $J$

From the resolved R branch lineshapes, we also extract lineshape parameters to facilitate fitting the unresolved Q branch. R branch lineshapes are well described by a saturated Voigt profile $f(\gamma,\sigma_D,\Delta,s_0)$ (Appendix~\ref{sec:lineshape}) parameterized by $s_0$, Doppler width $\sigma_{D}$, laser frequency detuning $\Delta$, and homogeneous width given by 
\begin{equation}
    \gamma\equiv n \sigma_{PB} \overline{v}_{rel}/\pi
\end{equation}
The pressure-broadening cross sections $\sigma_{PB}$ provide a crucial upper limit on the total inelastic cross sections~\cite{Baranger1958}, as well as a proxy for dephasing in our (incoherent) rate equation model (Appendix~\ref{sec:rate_eqn_model}). $\overline{v}_{rel}$ is the ensemble-averaged relative speed of the C$_{60}-$BG system, and takes typical values of several hundred m/s.  

Next, we consider the model for vibrational relaxation. The high vibrational density of states $\sim10^2/$cm$^{-1}$~\cite{Menendez1995, Tardy1968} for C$_{60}$ at the energy of the 1185~cm$^{-1}$ excited state prompts a $J$-independent ``reservoir model"~\cite{Preston2008,Feldman1973} for the first step of vibrational relaxation, in which $\sigma_{vib}(\textrm{reservoir},J_i) = \langle \sigma_{vib} \rangle$, with the thermal average $\langle ... \rangle$ defined similarly as in Equation~\ref{eq:sigma_rot}. In principle, after a sufficient number of inelastic collisions, population can return to the ground vibrational state and reenter the optical cycle, with vibrational cross section $\langle\sigma_{vib}'\rangle$. Finally, diffusion causes pumped molecules to be replaced by thermalized molecules at a rate A$_{diff}$.

\section{C$_{60}-$Ar collision cross sections} 
Having established this physical model for C$_{60}$ optical pumping and relaxation, we can now discuss the C$_{60}-$Ar collision cross sections extracted from fitting the Q and R branch saturated absorption profiles. First, we set $\langle\sigma_{vib}'\rangle= 0$, justified by the expectation that collisional relaxation of vibrations across the large 272~cm$^{-1}$ energy gap between the ground and first excited vibrational state~\cite{Menendez1995} is much slower than the diffusive transport through the cavity mode. We also calculate the elastic cross-section $\sigma_{el}$ as a function of collision energy using a semi-classical model (Appendix~\ref{sec:sigma_el}). The diffusion rate $A_{diff}$ can then be obtained from 
\begin{equation}\label{eq:A_diff1}
    A_{diff} = \frac{2 \overline{v}_{rel}}{n \overline{\sigma}_{tot} x_{eff}^2 }
\end{equation}
with (Appendix~\ref{sec:diff_const})
\begin{equation}
    \overline{\sigma}_{tot} = \overline{\sigma}_{el}+\langle\sigma_{rot}\rangle+\langle\sigma_{vib}\rangle
\end{equation} Intuitively, $x_{eff}$ represents the (ensemble-averaged) typical length scale over which a C$_{60}$ molecule must diffuse in order to leave the optical pumping region.

Next, $x_{eff}$, $\langle\sigma_{rot}\rangle$, and  $\langle\sigma_{vib}\rangle$ are varied to obtain the best simultaneous agreement of the rate equation model simulated profiles to the Q branch and R($J=170$) data as quantified by the reduced $\chi^2$ statistic. This procedure yields $x_{eff}=0.52\pm0.03 ~\textrm{mm}$,  $\langle\sigma_{rot}\rangle=123_{-36}^{+60}~\textrm{\AA}^2$, $\langle\sigma_{vib}\rangle=0.07\pm0.03~\textrm{\AA}^2$ (uncertainties represent $68\%$ confidence intervals). For typical operating conditions of 0.25~Torr argon, Equation~\ref{eq:A_diff1} then gives $A_{diff}\approx$ 2,500~s$^{-1}$.

\section{Varying C$_{60}$ collision partners}

The inelastic cross sections vary as a function of the collision partner. For example, at fixed temperature and pressure, lighter BGs are expected to feature less momentum transfer per collision. Also, unlike structureless atoms, molecular rotors can exchange rotational in addition to collisional angular momentum, which can be expected to increase $\langle\sigma_{rot}\rangle$.
Having completed a careful characterization of C$_{60}-$Ar collisions, we therefore turn to the collision cross sections with other closed-shell BG species: Ne, He, D$_{2}$, and H$_{2}$. For each of these species, saturated absorption profiles of C$_{60}$ in the Q and R($J=170$) branches have been measured at various BG pressures and then simultaneously fit to the rate equation model.

\begin{table}[htbp!]%[ht]
\caption{Pressure broadening cross sections $\sigma_{PB}$ and the temperatures at which they were measured.} % title of Table
\centering % used for centering table
\begin{tabular}{c c c} % centered columns (4 columns)
\hline\hline %inserts double horizontal lines
BG species & $\sigma_{PB}~ ({\text{\AA}}^2)$ & T~(K)\\ [0.5ex] % inserts table
%heading
\hline % inserts single horizontal line
Ar & 430(4) & 138(2) \\ % inserting body of the table
Ne & 310(4) & 138(2) \\
He & 163(8) & 159(5) \\
D$_2$ & 281(5)& 156(3)  \\
H$_2$ & 243(8)& 162(4)  \\ [1ex] % [1ex] adds vertical space
\hline %inserts single line
\end{tabular}
\label{table:sigma_PB} % is used to refer this table in the text
\end{table}

\begin{figure}[htbp!]\centering
	\includegraphics[width=\columnwidth,scale=1]{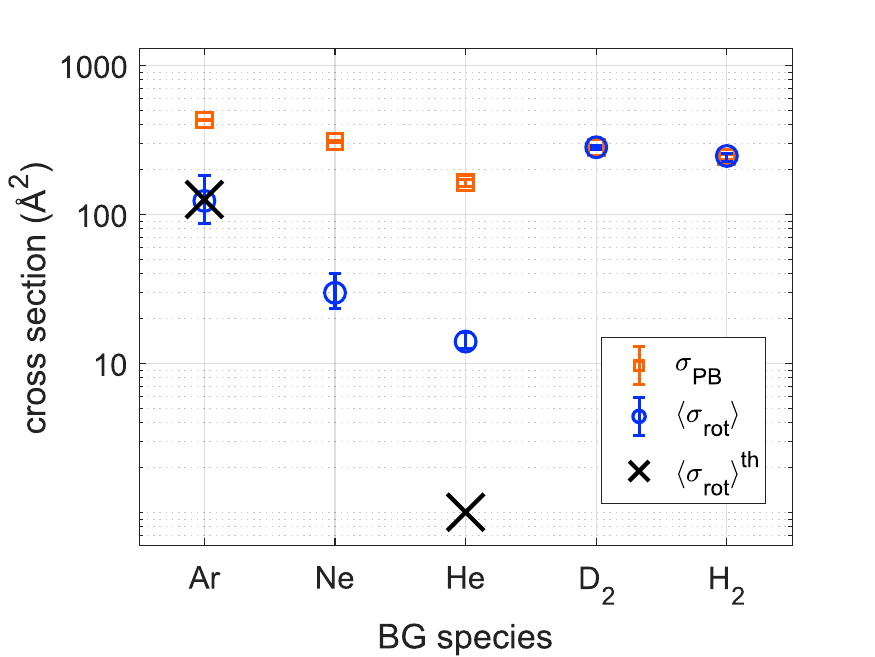}
	\caption{Average rotational inelastic collision cross sections $\langle \sigma_{rot}\rangle$ extracted from rate equation model fits to C$_{60}-$BG saturated absorption data. Total pressure-broadening cross section, $\sigma_{PB}$, obtained from R-branch fits, are plotted for reference. For rare gases with no rotational structure, $\langle\sigma_{rot}\rangle$ decreases with decreasing mass, reaching a minimum for He. By comparison, the presence of a rotational degree of freedom in D$_2$ and H$_2$ results in a drastic increase of $\langle \sigma_{rot}\rangle$ considering the small mass of both molecules. Error bars indicate $68\%$ confidence intervals and include the uncertainty on the effective diffusion length $x_{eff}$. Calculated values of $\langle \sigma_{rot}\rangle^{th}$ for C$_{60}-$Ar and C$_{60}-$He are plotted for comparison.}
	\label{fig:RG_fits}
\end{figure}

First, $\sigma_{PB}$ are straightforwardly obtained by fitting Voigt profiles to the resolved R-branch lineshapes and compiled in Table~\ref{table:sigma_PB}. We also fix $x_{eff}$ from the fits of the C$_{60}-$Ar system since it is not expected to vary with BG species. The resulting best-fit $\langle\sigma_{rot}\rangle$ values for each C$_{60}$-BG system are plotted in Fig.~\ref{fig:RG_fits}, together with the pressure broadening cross section $\sigma_{PB}$ determined from the R-branch data, shown for reference. The complete set of fitted profiles is shown in Appendix~\ref{sec:fitted_all}. Only Ar yields a measurable ($>10^{-2}\textrm{\AA}^2$) $\langle\sigma_{vib}\rangle$, consistent with the observation that the most efficient vibrational cooling, and hence the strongest absorption signal, is obtained with the most massive collision partner~\cite{Changala2019}. 
Two trends are immediately apparent: 1) for rare gas atoms with no rotational structure, $\langle\sigma_{rot}\rangle$ decreases with decreasing mass, reaching a minimum for He; and  2) for diatomic BGs with a rotational degree of freedom, $\langle\sigma_{rot}\rangle$ increases dramatically, despite having masses comparable to or smaller than He. The large values of $\langle \sigma_{rot}\rangle$ for C$_{60}-$D$_2$ and C$_{60}-$H$_2$ suggest that rotation-rotation energy transfer dominates~\cite{Taatjes1988}. Moreover, since $\langle\sigma_{rot}\rangle\approx\sigma_{PB}$, collisional dephasing is negligible in these two systems. These two observations are consistent with the expectation that molecular rotors can induce rotation-rotation relaxation from long range~\cite{Michael2021}, where the interaction potential changes slowly and dephasing is less significant. By contrast, the strikingly small value of $\langle\sigma_{rot}\rangle$ for the C$_{60}-$He interaction suggests that this complex is very ``slippery", similar to the situation observed for C$_{60}-$K~\cite{Rayane2000}.

\section{Comparisons with ab initio calculations}
\begin{figure}[htbp!]\centering
	\includegraphics[width=.7\columnwidth,scale=1]{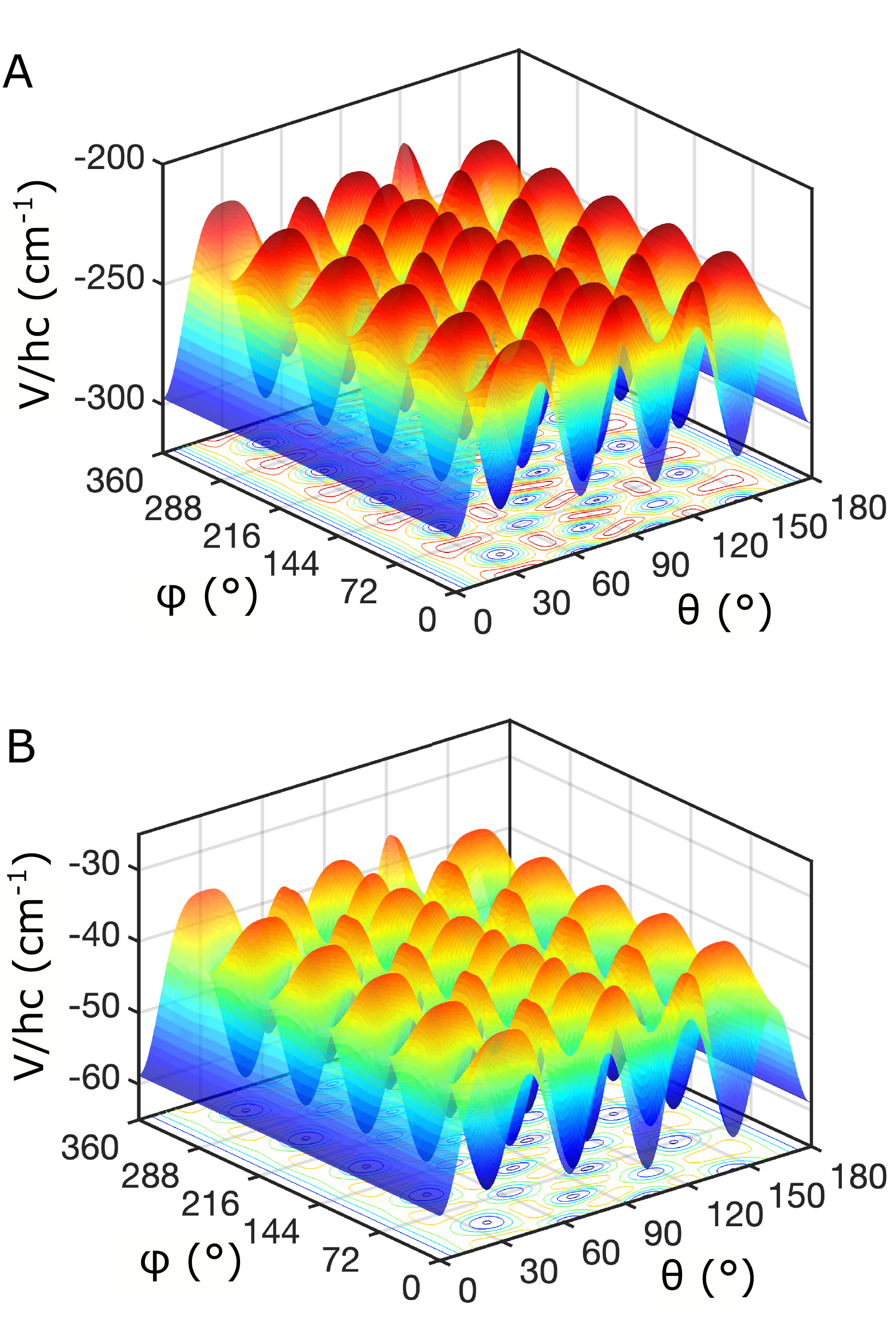}
	\caption{Calculated potential energy surfaces for C$_{60}-$BG interactions. A) Potential energy surface for C$_{60}-$Ar calculated at the equilibrium bond length of $7.2~\text{\AA}$. Polar $\theta$ and azimuthal $\phi$ angles are defined with respect to a Cartesian coordinate system with its $x$ and $z$ axes along a two-fold and five-fold symmetry axes of C$_{60}$, respectively. The potentials have minima when the Ar atom is located ``above" a five- or six- membered ring of C$_{60}$. Local maxima occur in between the minima. The potential depth at these maxima is 20\% to 30\% smaller than at the minima. B) Potential energy surface for C$_{60}-$He calculated at the equilibrium bond length of $7.0~\text{\AA}$. Note the change in vertical scale. The potential anisotropy for this complex is significantly reduced compared to that of C$_{60}-$Ar, and is reflected in the smaller value of its inelastic rotational cross section. } 
	\label{fig:PES_theory}
\end{figure} 

To understand the physical origin of these results, we calculate the ground-state potential energy surface of the C$_{60}-$Ar and C$_{60}-$He complexes as functions of the position of the Ar or He atom, using \textit{ab-initio} density functional theory (Appendix~\ref{sec:calculate_PES}). We plot in Fig.~\ref{fig:PES_theory} the potentials for C$_{60}-$Ar and C$_{60}-$He at their equilibrium bond lengths of $7.2~\text{\AA}$  and $7.0~\text{\AA}$, respectively. Of the two systems, the C$_{60}-$He complex shows a significantly reduced potential anisotropy due to the smaller polarizability of He compared to Ar. This weak anisotropy is reflected in the small magnitude of $\langle \sigma_{rot}\rangle$ for C$_{60}-$He collisions. Finally, we performed accurate calculations of the various anisotropic terms in these potential energy surfaces, obtaining averaged rotationally inelastic cross sections $\langle\sigma_{rot}\rangle^{th}$=126~$\textrm{\AA}^2$ for C$_{60}-$Ar and 1~$\textrm{\AA}^2$ for C$_{60}-$He~\cite{klos2022}. These are plotted alongside the experimental data in Fig.~\ref{fig:RG_fits} for comparison. Excellent agreement is obtained for C$_{60}-$Ar, while C$_{60}-$He features a significant discrepancy. In the latter case, $\langle \sigma_{rot}\rangle^{th}$ arises from the linear combination of many small terms of opposite signs in the anisotropic potential, which may lead to a large uncertainty. This suggests that our data are sufficiently precise to benchmark \textit{ab initio} calculations of this unique collision complex. 

\section{Summary}

In conclusion, a simple nonlinear absorption measurement yields remarkably detailed results. Specifically, nonlinear absorption lineshapes via laser optical pumping of single R-branch transitions versus collections of rotational states in the Q branch provide pressure-broadening and rotational and vibrational inelastic collision cross sections between C$_{60}$ and various buffer gas species. We find that $\langle \sigma_{rot}\rangle/\langle\sigma_{vib}\rangle\gtrsim10^3$ for C$_{60}-$Ar around 150~K, and that $\langle\sigma_{vib}\rangle$ is not measurable at our current sensitivity for the other, less massive BGs. The small vibrational quenching cross sections are in line with C$_{60}$ being a hard, rigid sphere~\cite{Osawa1996}. Internal structure of diatomic BG's increases $\langle \sigma_{rot}\rangle$ dramatically, likely due to rotation-rotation relaxation induced from long-range. Our data are consistent with an exponential gap law for rotational energy exchange and reveal the energy flow from the external interaction to the internal modes of C$_{60}$ for a range of collision partners. 
We find excellent agreement of C$_{60}-$Ar rotational inelastic cross sections with \textit{ab initio} calculations. 

Our experiments and calculations motivate and benchmark future theoretical work, and open a new avenue for probing collisional relaxation dynamics of an unprecedentedly large and symmetric molecule. Specific and precise state-to-state cross sections could be determined in future experiments where the effect of a cw optical pump is probed with a frequency comb across all relevant rotational states simultaneously.

\begin{acknowledgments}
The authors thank D. Rosenberg and T. Q. Bui for technical assistance and discussions. This research was supported by AFOSR grant no. FA9550-19-1-0148, the
National Science Foundation Quantum Leap Challenge Institutes
Office of Multidisciplinary Activities (grant 2016244); the National
Science Foundation (grant Phys-1734006); the Department of Energy (DE-FG02-09ER16021); and the National
Institute of Standards and Technology. J.T. was supported by the Lindemann Trust in the form of a Postdoctoral Fellowship. J.K. and S.K. acknowledge support from the U.S. AFOSR grant no. FA9550-21-1-0153 and the NSF grant no. PHY-1908634.
\end{acknowledgments}

\appendix
\section{Experimental Details}
The high finesse spectroscopy cavity (finesse $F=12,000$; Gaussian beam waist $w_0=1.1~$mm) intersects a $6\times6\times6$~cm$^3$ aluminum cold cell held between 130-160~K. C$_{60}$ evaporates from an oven and is entrained in a flow of buffer gas (BG) into the cell, where it thermalizes to the BG and cell temperature~\cite{Changala2019}.

\begin{figure}[htbp!]
    \centering
    \includegraphics[width=\columnwidth]{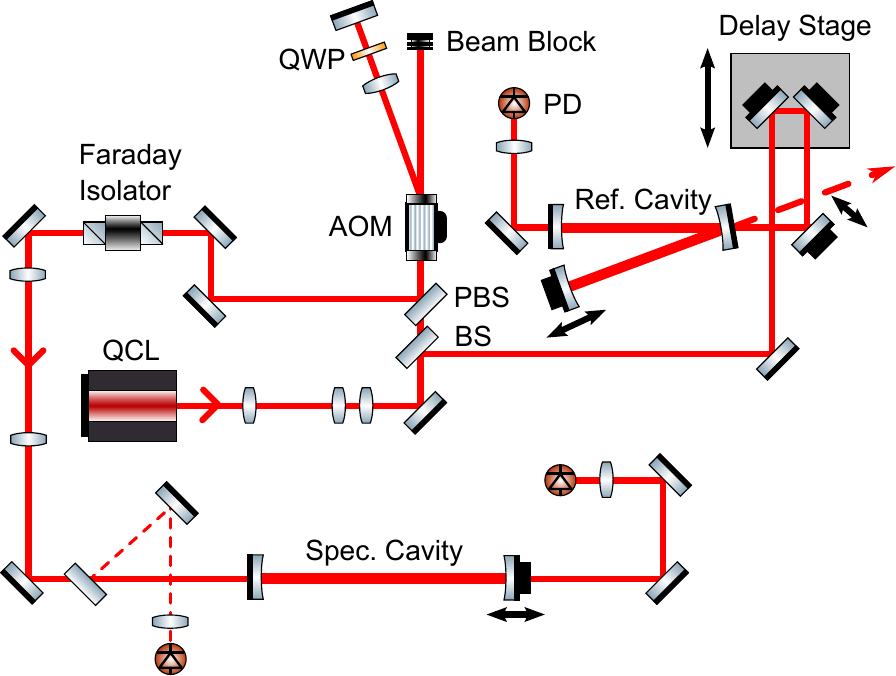}
    \caption{Optical beam path for the experiment. The 8.4$~\mu$m beam from the QCL is split into two paths by a beamsplitter (BS): 1) The reflected beam is locked to a reference cavity which provides optical feedback to narrow the laser linewidth (see text). To enable optical feedback stabilization, a delay stage and bullet mount piezo mirror~\cite{Briles2010} provide slow and fast control of round-trip optical phase, while a piezo-mounted cavity mirror stretches the reference cavity length. 2) The transmitted beam is PDH locked to the spectroscopy cavity, with a double-pass AOM and piezo mounted cavity mirror for the fast and slow frequency feedback, respectively. The absorption signal is monitored by a photodiode on the cavity transmitted output. Double-sided arrows depict direction of travel of mirrors. PBS: polarizing beamsplitter; QWP: quarter waveplate; PD: photodetector.}
    \label{fig:setup}
\end{figure}

The optical beam path is shown in Fig.~\ref{fig:setup}. The light is derived from an Alpes Lasers continuous wave quantum cascade laser (cw-QCL) powered by a home-built current driver. To narrow down the 20~MHz free-running linewidth for efficient coupling into the spectroscopy cavity, half of the beam power is picked off and frequency locked to an external three-mirror v-shaped reference cavity machined out of a single block of aluminum and purged with N$_2$ gas (free spectral range = 725~MHz). The cavity length can be tuned over twice the free spectral range by a ring piezo mounted to the cavity spacer. Only light resonant with the cavity is reflected back to the QCL, providing reduction of the laser linewidth via optical feedback stabilization~\cite{Laurent1989,Maisons2010,Zhao2021}. While the relative laser-reference cavity linewidth is estimated to be $<1~$kHz, jitter of the reference cavity itself widens the absolute laser linewidth to $\lesssim10$~kHz.  The cavity frequency is passively stable to within a few ppm (a few kHz) over an hour, whereas the optical feedback stabilization locking bandwidth is 200~MHz. However, the laser-cavity round-trip optical phase requires active stabilization due to the $\sim1$~m path length. We dither the free-running laser frequency by applying a 100~kHz sinewave on top of the QCL drive current and demodulate the reference cavity transmitted intensity to produce an error signal for the round-trip phase. This is fed back to the delay stage and bullet-mounted piezo mirror that provide slow and fast control, respectively, of the path length to maintain optimal conditions for optical feedback stabilization. To scan the optical frequency of the reference cavity mode, the cavity length and round-trip phase are tuned simultaneously. The stabilized QCL beam is locked to the spectroscopy cavity using a Pound-Drever-Hall (PDH) lock, with a piezo-mounted cavity mirror and double-pass acousto-optic modulator (AOM) providing slow and fast frequency control, respectively.

To obtain an absorption spectrum, the spectroscopy and reference cavity lengths and path length delay are scanned simultaneously with the cw-QCL current back and forth over a frequency range of 500-600~MHz over a 4-s period while the transmission through the spectroscopy cavity is monitored with a liquid-nitrogen cooled detector (Kolmar KLD-0.5-J1/11/DC). A separate 1064~nm laser (Coherent Mephisto) reflected off the reference cavity scanning mirror monitors the relative frequency excursion by counting interference fringes. 

To determine the cell pressure, a long 1/4" diameter stainless steel bellows was attached to the cold cell, and the other end to a capacitance manometer thermally anchored to the chamber wall at room temperature. The zero-flow condition ensures the pressure measurement is consistent with that at the cold cell, while the capacitance manometer operates at room temperature. We assume the pressure is essentially that of the BG alone, as the concentration of C$_{60}$ is estimated at the ppm level (Section~\ref{sec:c60_c60_collisions}). The cold cell temperature is measured with a thermocouple attached to the cell wall. Pressure in the cold cell could be varied from 100-500~mTorr by choking the pumping speed and varying the BG flow rate.

The intracavity power is obtained by dividing transmitted power by the mirror transmission of 200~ppm, independently verified by cavity finesse and transmission measurements. The intensity is obtained by dividing by $\pi w_0^2$ where $w_0=1.1~$mm is the Gaussian beam waist obtained from ABCD calculations. 

Residual etalon fringes in our optical setup have a period of $\sim200~$MHz. Since this is much larger than the linewidth of the resolved R-branch peaks, etaloning can be removed by fitting to a suitable lineshape function (Section~\ref{sec:lineshape}). However, the Q-branch contour spans $\sim 270~$MHz, completely masking the etalon fringes and making them a dangerous source of systematics. We carefully tilted and isolated optical surfaces to suppress the intensity fluctuations due to etalons down to about $0.2\%$. These residual uncertainties are included in the Q-branch data as gray error bands. 

\section{Effect of C$_{60}-$C$_{60}$ collisions.}\label{sec:c60_c60_collisions}
Here we estimate the C$_{60}-$C$_{60}$ collision rate. First, we estimate the number density at typical operating conditions of 150~K and 300~mTorr buffer gas pressure. The integrated Napierian absorbance of the unsaturated Q branch was measured to be $\int \textrm{ln}(I_0/I(\nu))d\nu =9 \times10^{-4}~\textrm{cm}^{-1}$. The total  band absorbance for all C$_{60}$ isotopologues is therefore estimated to be 

\begin{align}
    A &= 3\times\frac{1}{0.51}\times\int \textrm{ln}(I_0/I(\nu))d\nu\\ 
    &= 5.3 \times10^{-3}~\textrm{cm}^{-1}
\end{align} 
The factor of 3 is because the Q branch accounts for only one third of the total vibrational band strength, and the factor of $1/0.51$ is because $^{12}$C$_{60}$ only accounts for 51$\%$ of all C$_{60}$ isotopologues~\cite{Harter1992}. 

From the molar absorptivity $\Psi = \textrm{ln}(10)\times9.9~$km/mol~\cite{Iglesias-Groth2011a}, cavity finesse $F=12,000$, and buffer gas cell length $l=6$~cm, we can estimate the number density in the cell

\begin{align}
n &=\frac{A}{2 \Psi  F l/\pi} \\
& = 5\times10^{-11}~\textrm{mol/L}
\end{align}

\begin{table}[htbp!]%[ht]
\caption{Estimate of relative C$_{60}$-C$_{60}$ collision rate} % title of Table
\centering % used for centering table
\begin{tabular}{c c } % centered columns (4 columns)
\hline\hline %inserts double horizontal lines
parameter & C$_{60}$-C$_{60}$ : C$_{60}$-BG \\ [0.5ex] % inserts table
%heading
\hline % inserts single horizontal line
$n$ & $1/(6 \times 10^{5})$ \\ % inserting body of the table
$\overline{\sigma}_{tot}$ & 4  \\
$\overline{v}_{rel}$ & 1/13 (H$_2$) - 1/3 (Ar)  \\
$\Gamma_{tot}=n \overline{\sigma}_{tot} \overline{v}_{rel}$ & $5\times 10^{-7} - 2\times 10^{-6}$ \\ [1ex] % [1ex] adds vertical space
\hline %inserts single line
\end{tabular}
\label{table:C60_C60_collision_rate} % is used to refer this table in the text
\end{table}
Note that this provides an underestimate of C$_{60}$-C$_{60}$ collision rate since it is only based on the measured absorbance from cold, ground vibrational state C$_{60}$.
By contrast, at 300~mTorr and 150~K, the buffer gas number density is $3\times10^{-5}~\textrm{mol/L}$, $6\times10^{5}$ higher than that calculated for C$_{60}$. The total collision rate is therefore expected to be between $0.5-2\times10^{-6}$ times less for C$_{60}$-C$_{60}$ collisions than for C$_{60}$-BG collisions (Table~\ref{table:C60_C60_collision_rate}). The minimum inelastic cross sections we can measure are of the order $10^{-5}\times \overline{\sigma}_{tot}$. Hence, even in the extreme case $\sigma_{inel}=\overline{\sigma}_{tot}$ for C$_{60}$-C$_{60}$ collisions, relaxation from C$_{60}$-C$_{60}$ collisions would not be detected by our experiment, and we ignore them in this work.

\section{Langevin diffusion constant.}\label{sec:diff_const}
The drag force from C$_{60}$ colliding with BG atoms is given by
\begin{align}
    \langle F \rangle_t &= -\alpha \langle v \rangle_t\\
    &=\langle \Delta p\rangle_t / \tau\\
    &=-2 \mu \langle v\rangle_t /\tau\\
\end{align}
where $\alpha$ is the drag force coefficient, $\langle v\rangle_t$ is the speed of C$_{60}$, $\langle \Delta p\rangle_t$ is the momentum change experienced by C$_{60}$ per collision,  $\tau$  is the mean time between C$_{60}$-BG collisions, and $\mu$ is the reduced mass of the C$_{60}$-BG system. The symbol $\langle \cdots\rangle_t$ is the ``long-time average" where the averaging time is large compared to $\tau$. Therefore, the drag force coefficient is given by
\begin{equation}
    \alpha = 2\mu / \tau
\end{equation}

The spatial diffusion coefficient is given by~\cite{Pathria2011,Zwanzig2001,Hodapp1995}
\begin{align}
    D &= \frac{kT}{\alpha}\\
    &= k T \frac{\tau}{2 \mu}\\
    &=\frac{1}{2}\frac{\overline{v}_{rel}}{n \overline{\sigma}_{tot}}\\
\end{align}

where $n$ is the number density of the buffer gas, $\overline{v}_{rel}$ is the ensemble averaged relative speed of the C$_{60}$-BG system, and $\overline{\sigma}_{tot} = \sigma_{el}(\overline{v}_{rel})+\langle\sigma_{rot}\rangle+\langle\sigma_{vib}\rangle$ is approximately the thermal and $J-$ averaged total cross section. Here we are implicitly making the approximation for the elastic collision rate constant $\overline{\sigma_{el} v_{rel}}\approx\sigma_{el}(\overline{v}_{rel})\overline{v}_{rel}$, which has an error of about 2\% for our temperatures of interest. Elastic cross sections $\sigma_{el}$ as a function of collision energy are calculated in Section~\ref{sec:sigma_el}.

The diffusion time $t_{diff}$, mean squared distance $ \langle \Delta x^2 \rangle$, and diffusion coefficient $D$ are related by
\begin{equation}
    2 N D~ t_{diff} = \langle \Delta x^2 \rangle,
\end{equation}
where $N$ is the dimensionality~\cite{Pathria2011}. Here, $N = 2$ because we are only sensitive to diffusion in the two spatial dimensions transverse to the cavity mode (which is very nearly collimated). Setting the diffusion rate $A_{diff} = (t_{diff})^{-1}$ and the effective diffusion length $x_{eff}=\sqrt{ \langle \Delta x^2 \rangle} $ yields

\begin{equation}\label{eq:A_diff}
    A_{diff} = \frac{2 \overline{v}_{rel}}{n \overline{\sigma}_{tot} x_{eff}^2 }
\end{equation}

\section{Calculating elastic cross sections.}\label{sec:sigma_el}

We first estimate the long-range van-der-Waals dispersion coefficient $C_6$ for the C$_{60}-$BG systems. First, we  calculate the dynamic dipole polarizability tensor $\alpha_{ij}(\omega)$ with ${i,j=x}$, $y$, and $z$
as function of imaginary-valued frequency $\omega$ for rigid C$_{60}$ using the coupled-cluster propagator with single- and double-excitations and the STO-3G basis set in Molpro \cite{Molpro}. 
The  body-fixed coordinate system $(x, y, z)$ is  defined as in Fig.~\ref{fig:PES_theory} of the main text.
The basis set in this calculation, however, is not sufficiently accurate. We therefore 
also calculate the static, $\omega=0$ dipole polarizability tensor  in Q-Chem \cite{qchem}  with the much-larger 6-311G(d,p) basis and the functionals of Refs.~\cite{Becke88,Perdew86}.   

\begin{table}[htbp!]
\caption{Static dipole polarizability  tensor $\alpha_{ij}(0)$ of C$_{60}$ in units of $a_0^3$.}
\centering
\label{tab:alfa}
\begin{tabular}{c|r@{.}l|r@{.}l|r@{.}l|}
&\multicolumn{2}{c|}{$x$} & \multicolumn{2}{c|}{$y$}  & \multicolumn{2}{c|}{$z$}  \\
\hline
$x$ &502 & 6413 &0 & 0000 & 0 & 01262\\
$y$ &0 &0000 & 502 &6394 &0 & 0000\\
$z$&0 &01262 &0 & 0000& 502 & 6453
\end{tabular}
\end{table}

\begin{figure}[htbp!]
\centering
	\includegraphics[scale=0.4,trim=40 20 40 80,clip]{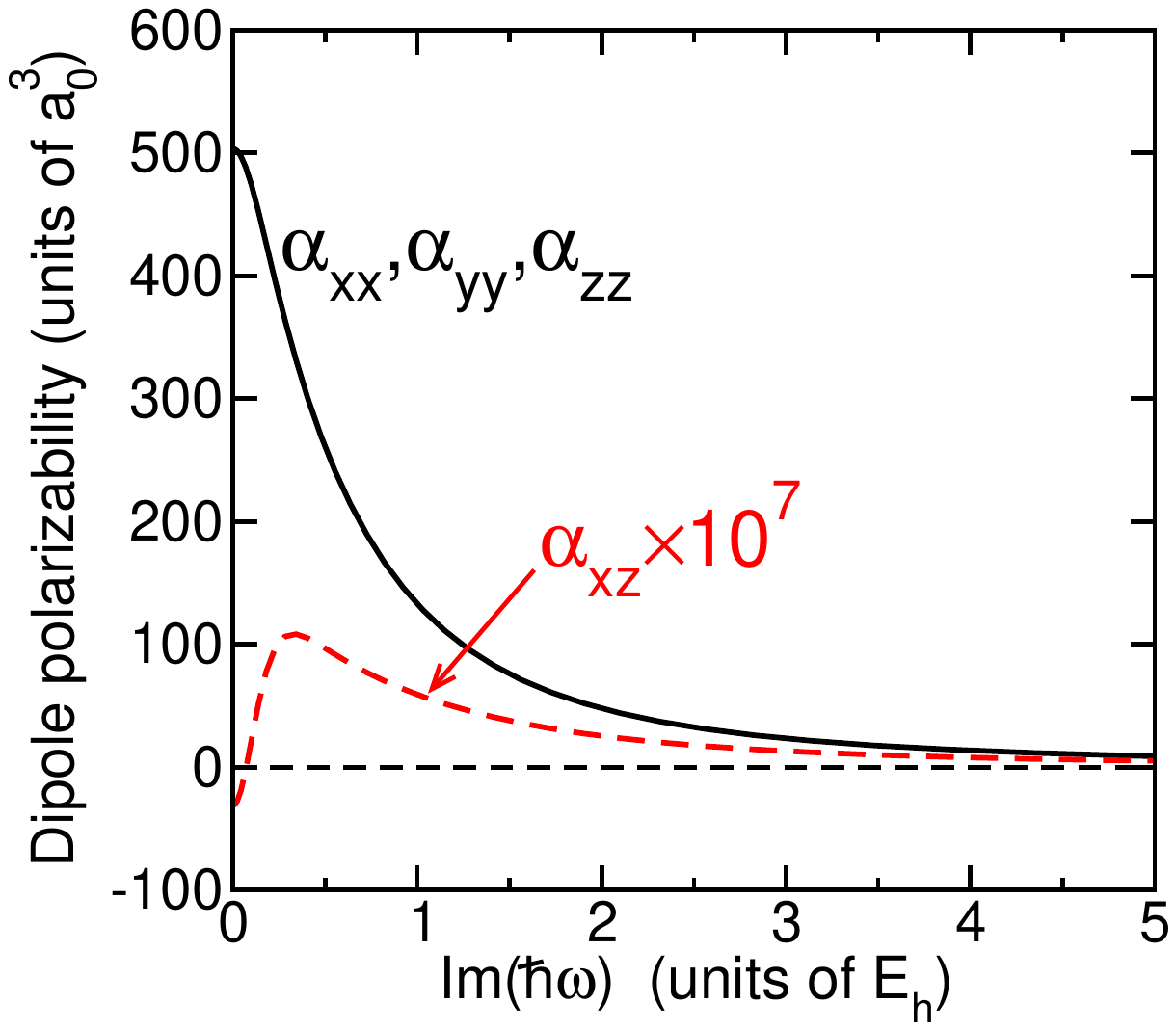}
	\caption{Our calculated   dipole polarizability tensor elements of  C$_{60}$ as  functions of imaginary frequency ${\rm Im}(\omega)$. The  off-diagonal  tensor elements have been multiplied by $10^7$ for clarity.}
	\label{fig:DipPol}
\end{figure}

These accurate  values of the static dipole polarizability tensor are shown in Table~\ref{tab:alfa}. The corresponding isotropic static polarizability of  C$_{60}$ is $\alpha_{\rm iso}(0)=\sum_i \alpha_{ii}(0)/3=502.6~a_0^3$  or 74.54 \AA$^3$, where $a_0$ is the Bohr radius,
and agrees well with the measured static polarizability of $79\pm 4$ \AA$^3$ \cite{Ballard2000}. 
The anisotropic components of $\alpha_{ij}(0)$ are six orders of magnitude smaller. 
Finally, we uniformly scale the  coupled-cluster calculations of the  dynamic dipole polarizability tensor such
that its  isotropic  static dipole polarizability coincides with that from the  Q-Chem calculation.
The resulting dipole polarizability tensor as functions of imaginary frequency are shown in Fig.~\ref{fig:DipPol}.

\begin{table}[htbp!]
\caption{$C_6$ coefficients of C$_{60}-$BG in units of $E_h a_0^6$.}
\centering
\label{tab:C6_coeffs}
\begin{tabular}{c|r@{.}l|}
&\multicolumn{2}{c|}{C$_6$ [$E_h a_0^6$]}   \\
\hline
C$_{60}-$He &369 & 54452 \\
C$_{60}-$Ne &749 &39667 \\
C$_{60}-$Ar &2523 &00369 \\
C$_{60}-$D$_2$,H$_2$ &1094 &61595 \\
\end{tabular}
\end{table}

By combining our frequency-dependent polarizability tensor of \ C$_{60}$  and the frequency-dependent dipole polarizability of the noble gas atoms from Ref.~\cite{DEREVIANKO2010323} in the Casimir-Polder formula, we find the isotropic $C_{6}$ coefficients listed in Table~\ref{tab:C6_coeffs}. Here, $E_{\rm h}$ is the Hartree energy. We are in reasonable agreement with the  $2035~E_{\rm h}a_0^6$ value for C$_{60}-$Ar from  Ref.~\cite{Han1995}.  Since we only consider electronic contribution to the polarizability, the $C_6$ coefficients for C$_{60}-$D$_2$ and C$_{60}-$H$_2$ are identical.

Finally, the elastic cross sections as a function of the relative collision velocity $v_{rel}$ in the approximation of high-energy phase shift and $R^{-6}$ long-range potential can be computed from~\cite{Busch1966,Child} 
\begin{equation}
 \sigma_{el} = 8.083 \left(\frac{C_6}{\hbar v_{rel}}\right)^{2/5}   
\end{equation}

\begin{figure}[htbp!]
\centering
	\includegraphics[width=\columnwidth]{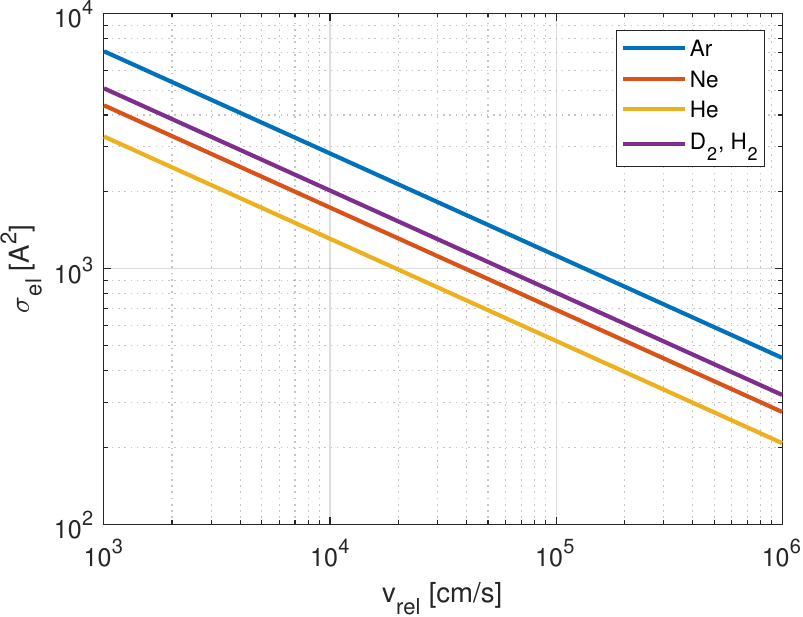}
	\caption{Our calculated semi-classical elastic cross sections $\sigma_{el}$ as functions of relative collision velocity $v_{rel}$.}
	\label{fig:calc_sigma_El}
\end{figure}

 These are plotted in Fig.~\ref{fig:calc_sigma_El} for all BG species considered in this work.

\section{Rate equation model.}\label{sec:rate_eqn_model}
The small rotational constant, rapidly increasing vibrational density of states with energy, and the thousands of accessible rotational states even at cryogenic temperatures, render the full theoretical description of collisional relaxation dynamics a formidable task. We turn to a rate equation model to extract the most salient physics, the partitioning of collisional energy into inelastic vibrational and rotational transitions.  

In our cw spectroscopy scans, the maximum frequency slew rate is 400~MHz/s $\approx$ 40$\times\gamma$/s. 1/40 seconds is by far the longest timescale among the mean time between collisions ($\lesssim10^{-7}$~s) or diffusion-limited beam transit time ($\lesssim10^{-3}$~s). Also, the translational and rotational degrees of freedom are assumed to be completely thermalized to the BG and cell wall. This is supported by the rotational Boltzmann spectrum observed in our previous work~\cite{Changala2019}.

We therefore assume the probed system is always in steady state. Calculating the absorption cross section then amounts to finding the steady state rovibrational populations in the presence of optical pumping, collisions, and diffusion. We solve a steady-state rate equation of the form 

\begin{align}\label{eq:rate_eqn}
    \frac{d}{dt}\mathbf{ v_0} &= \mathbf{(K+R) v_0} = 0,
\end{align}
where $\mathbf{v_0}$ is the vector containing steady-state populations of every rovibrational state under consideration. $\mathbf{K}$ is the transition rate matrix for all thermal (i.e., collision and diffusion) processes and $\mathbf{R}$ is that for optical pumping. 
$\mathbf{(K+R) v_0} $ in block-matrix form is
\begin{widetext}

\begin{equation}\label{eq:rate matrix structure}
\left(
\begin{array}{c|c|c}
\begin{matrix}\\ v = 0 \\ \\\end{matrix} & v = 0\leftarrow v=1 & v = 0 \leftarrow \textrm{res}\\
\hline
\begin{matrix}\\ v = 0\rightarrow v=1   \\ \\\end{matrix}& v = 1 & v=1\leftarrow\textrm{res} \\
\hline
\begin{matrix}\\ v = 0\rightarrow\textrm{res} \\ \\\end{matrix}& v=1\rightarrow\textrm{res} & \textrm{res} \\
\end{array}
\right)
\begin{pmatrix}
    \\
    \mathbf{N_0} \\
    \\
    \hline \\
    \mathbf{N_1} \\
    \\ 
    \hline
    \\
    \mathbf{N_{res}},
    \\ \\
    
\end{pmatrix}
\end{equation}
\end{widetext}

where 
\begin{equation}
\mathbf{N_0} = \begin{pmatrix}
    P^J_{v=0}(J=0),P^J_{v=0}(J=1),\cdots,P^J_{v=0}(J=J_{end})
\end{pmatrix}
\end{equation}

is the $N_0\times 1$ vector of populations of each $J$-state in the $v=0$ manifold (where $N_0=J_{end} + 1$), 
\begin{widetext}

\begin{align}
    \mathbf{N_1} = \bigg(&P_{v=1}^{J+1}(J=0),&P_{v=1}^{J+1}(J=1),&~~~~\cdots,&P_{v=1}^{J+1}(J=J_{end}),\\
    &P_{v=1}^{J}(J=0),&P_{v=1}^{J}(J=1),&~~~~\cdots,&P_{\nu=1}^{J}(J=J_{end}),\\
    &P_{v=1}^{J-1}(J=0),&P_{v=1}^{J-1}(J=1),&~~~~\cdots,&P_{v=1}^{J-1}(J=J_{end})\bigg)
\end{align}
\end{widetext}

is the $N_1\times 1$ vector of populations of each $J$-state in the $v=1$ manifold (where $N_1=3\times (J_{end}+1)$), and $\mathbf{N_{res}}$ is the $1\times 1$ entry corresponding to the total population residing in the dark vibrational reservoir. Here, $P_v^R(J)$ denotes the population in the quantum state $|R,J,v\rangle$, where the ``pure rotational" angular momentum quantum number $R$ arises from the coupling of $J$ to the vibrational angular momentum due to Coriolis forces~\cite{Changala2019}. In the $v=1$ vibrational excited state, $R=J,|J\pm 1|$  thereby sorting the excited state into three Coriolis manifolds. We account for $J$ up to $J_{end}=470$. We have assumed the lab-fixed and body-fixed projections, $m$ and $k$, respectively, are mixed infinitely quickly which corresponds to instantaneous and uniform reorientation of the C$_{60}$ molecule. The steady-state solution equals the zero-eigenvalue eigenvector of the total transition rate matrix $ \mathbf{K+R}$, which yields the steady-state $\mathbf{N_0} $, $\mathbf{N_1} $, and $\mathbf{N_{res}} $.

The relaxation matrix $\mathbf{K}$ is the sum of rate matrices for all thermal processes: rotationally inelastic collisions, vibrationally inelastic collisions, diffusion of molecules out of the optical pumping beam, and spontaneous emission/black body pumping,

\begin{equation}
     \mathbf{K} = n \overline{v}_{rel} \left( \mathbf{K_R} +  \mathbf{K_V} \right) + A_{diff} \mathbf{K_D} + \mathbf{K_{SE}},
\end{equation}
where the diffusion rate $A_{diff}$ is derived in Section~\ref{sec:diff_const}. The rotational relaxation matrix $\mathbf{K_R}$ only couples $J$ states within a vibrational state, i.e.
\begin{equation} \label{eq:KR structure}
\mathbf{K_R}=\sigma_{rot}\left(\begin{array}{c|c|c}
  \begin{matrix}
    \\
    \mathbf{K_R^{(1)}}
  \\ \\
  \end{matrix}
  & 0   & \begin{matrix}0 \\ \vdots \\0 \\\end{matrix}\\
\hline 
  \begin{matrix}
  \\
  0
  \\ \\
  \end{matrix} 
  &  \mathbf{K_R^{(2)}} & \begin{matrix}0 \\ \vdots \\0 \\\end{matrix}\\
\hline
0 \cdots 0 & 0 \cdots 0 & 0
\end{array}\right)
\end{equation}
The upper triangle, not including the main diagonal, corresponds to downward transitions (i.e., that decrease rotational energy),
\begin{equation}
    \{\mathbf{K_R^{(1)}}\}_{f,i>f} = g_f ~\textrm{exp}(-\Delta E_{f,i}/kT),
\end{equation}
where $f,i$ index the final and initial states, respectively, and $g_f$ and $\Delta E_{f,i}$ are defined in the main text. By detailed balance, the lower triangle is given by
\begin{equation}
    \{\mathbf{K_R^{(1)}}\}_{f,i<f} =\frac{g_f}{g_i} \textrm{e}^{-(E_f-E_i)/kT}\{(\mathbf{K_R^{(1)}})^\intercal\}_{f,i<f}
\end{equation}
Finally, to conserve population, the diagonal elements $\{\mathbf{K_R^{(1)}}\}_{f,i=f}$ are chosen to make the columns sum to zero.

We have assumed the different Coriolis manifolds are uncoupled by rotational relaxation. Hence, $\mathbf{K_R^{(2)}}$ is block-diagonal

\begin{equation} 
 \mathbf{K_R^{(2)}} = \left(\begin{array}{c|c|c}
  \begin{matrix}
    \\
   \mathbf{K_{R(l=-1)}^{(2)}}
  \\ \\
  \end{matrix}
  & 0   & 0 \\
\hline 
  \begin{matrix}
  \\
  0
  \\ \\
  \end{matrix} 
  & \mathbf{K_{R(l=0)}^{(2)}}& 0 \\
\hline
 \begin{matrix}
  \\
  0
  \\ \\
  \end{matrix} & 0 &\mathbf{K_{R(l=+1)}^{(2)}}
\end{array}\right)
\end{equation}
with the $\mathbf{K_{R(l=-1,0,+1)}^{(2)}}$ matrices each defined similarly to $\mathbf{K_{R}^{(1)}}$.

The vibrational relaxation matrix $\mathbf{K_V}$ couples the vibrational manifolds to the vibrational reservoir. Direct $v=1\longleftrightarrow v=0$ transitions due to inelastic collisions are neglected:

\begin{equation}\label{eq:KV structure}
\mathbf{K_V}=\left(
\begin{array}{c|c|c}
\begin{matrix}\\ \mathbf{D^{(1)}} \\ \\\end{matrix} & 0 &  \mathbf{K_{V}^{( 0 \leftarrow \textrm{res})}}\\
\hline
\begin{matrix}\\ 0  \\ \\\end{matrix}& \mathbf{D^{(2)}} & \mathbf{K_{V}^{( 1 \leftarrow \textrm{res})}} \\
\hline
\begin{matrix}\\\mathbf{K_{V}^{( 0\rightarrow\textrm{res} )}}\\ \\\end{matrix}& \mathbf{K_{V}^{( 1\rightarrow\textrm{res} )}} & \mathbf{D^{(3)}} \\
\end{array}
\right)
\end{equation}
where
\begin{align}
     \mathbf{K_{V}^{(1\rightarrow\textrm{res} )}}&= \sigma_{vib}\begin{pmatrix}
         1,1,\cdots,1
     \end{pmatrix}\\
     \mathbf{K_{V}^{(1\leftarrow\textrm{res} )}}&= \left(\mathbf{K_{V}^{(1\rightarrow\textrm{res} )}}\right)^{\intercal}\odot  \left(\frac{Z_{v=1}}{Z_{res}}\times \left(\mathbf{P_f'}/\sum_f P_f'\right) \right)\\
     \mathbf{K_{V}^{(0\rightarrow\textrm{res} )}}&= \sigma_{vib}'\begin{pmatrix}
         1,1,\cdots,1
     \end{pmatrix}\\
     \mathbf{K_{V}^{(0\leftarrow\textrm{res} )}}&= \left(\mathbf{K_{V}^{(0\rightarrow\textrm{res} )}}\right)^{\intercal}\odot  \left(\frac{Z_{v=0}}{Z_{res}}\times \left(\mathbf{P_f''}/\sum_f P_f''\right) \right)\\
\end{align}
Here, $\odot$ denotes elementwise multiplication and the matrix of Maxwell-Boltzmann probability weights is given by
\begin{align}
         \mathbf{P_f''}&=\begin{Bmatrix}
    g_f\times e^{-E_f/kT}
    \end{Bmatrix}_f , f = 1,2,\cdots,N_0\\
 \mathbf{P_f'}&=\begin{Bmatrix}
    g_f\times e^{-E_f/kT}
    \end{Bmatrix}_f , f = N_0+1,N_0+2,\cdots,N_0+N_1
\end{align}
with $f$ denoting the final state index. Vibrational partition functions $Z_{\nu=0}$, $Z_{\nu=1}$, $Z_{res}$ are calculated in Section~\ref{ssec:vib partition function}. The block matrices on the main diagonal $\mathbf{D^{(1)}}$, $\mathbf{D^{(2)}}$, and  $\mathbf{D^{(3)}}$ are diagonal matrices, whose elements are set by making the columns of $\mathbf{K_V}$ sum to zero, as before.

The diffusion matrix $\mathbf{K_D}$ is constructed by filling each column with the Boltzmann probability weights of the final state:
\begin{align}
    &\frac{1}{Z_{vib}}\times\left(\left(Z_{v=0}+Z_{v=1}\right)~\mathbf{P_f}/\sum_f P_f~,~Z_{res}\right)^\intercal\\
\end{align}
where $Z_{vib}=Z_{v=0}+Z_{v=1}+Z_{res}$ and the Boltzmann weights are given by 
\begin{align}
 \mathbf{P_f}&=\begin{Bmatrix}
    g_f\times e^{-E_f/kT}
    \end{Bmatrix}_f &f = 1,2,\cdots,N_0+N_1
\end{align}    
Then, the diagonals of $\mathbf{K_D}$ are adjusted to make the columns sum to zero. Thus, any molecule leaving the pumping volume is replaced by a molecule drawn from the Boltzmann distribution.

Blackbody pumping and stimulated emission are described by the matrix $\mathbf{K_{SE}}$. For the spherical top, blackbody radiation and spontaneous emission couple states in $v=1\longleftrightarrow v=0$ with identical $R$ quantum number (we do not consider any radiative coupling with the vibrational reservoir). The upper triangular (i.e., downward transition) matrix elements contains terms for emission stimulated by blackbody radiation and spontaneous emission.  Here, $A_{21}^v = 1.4$~s$^{-1}$~\cite{Hilborn1982} is the vibrational Einstein A coefficient, calculated using an integrated band strength of 24 km mol$^{-1}$ from KBr matrix measurements~\cite{Iglesias-Groth2011a}. The total thermal radiative decay rate $B_{21}^v\rho(\nu)+A_{21}^v \approx A_{21}^v$ since, at $T=150~$K and $\nu/c=1185~$cm$^{-1}$~\cite{Demtroder2013}, 
\begin{align}
    B_{21}^v \rho(\nu) &= A_{21}^v/(\textrm{exp}(h\nu/kT)-1) \\
    &=\frac{A_{21}^v}{ 8.5\times10^{4}}
\end{align}
Furthermore, the probed vibrational bandwidth is less than 7~cm$^{-1}$. The variation in the A-constant (which scales as $\nu^3$) is only 2$\%$ across this entire range, so we assume it to be $J$-independent. 

With these approximations, we now define  upper triangular matrix elements of $\mathbf{K_{SE}}$:
\begin{equation}
    \{\mathbf{K_{SE}}\}_{f,i>f} =\begin{cases} 
      A_{21}^v & R_i = R_f \textrm{~(dipole selection rule)}\\
      0 & \textrm{otherwise} 
   \end{cases}
\end{equation}
The lower triangular matrix elements are determined by detailed balance, i.e.,
\begin{equation}
    \{\mathbf{K_{SE}}\}_{f,i<f} = \frac{g_f}{g_i} \textrm{e}^{-(E_f-E_i)/kT}\{(\mathbf{K_{SE}})^\intercal\}_{f,i<f} 
\end{equation}
Then the diagonals are again chosen to make the columns sum to zero. 

The pumping matrix accounts for homogeneous collisional (dephasing and lifetime) and inhomogeneous (Doppler) broadening and pumping from counterpropagating beams. We first consider pumping of a single velocity class with Doppler shift $\delta$ in the lab frame, with the pump laser frequency at $\nu_l$. 

\begin{figure}[htbp!]
    \centering
    \includegraphics[width=.7\columnwidth]{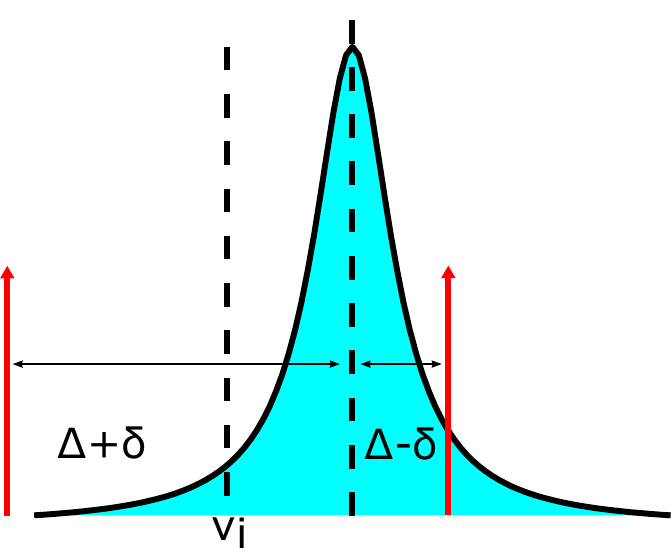}
    \caption{Pumping of a single velocity class (Doppler shifted to $\nu_i+\delta$)  by counterpropagating beams with lab-frame detuning $\Delta$.}
    \label{fig:SI_pumping_model}
\end{figure}

Off-diagonal matrix elements of $\mathbf{R_{OP}}$ are given by the photon absorption and stimulated emission rates:
\begin{equation}
    \{\mathbf{R_{OP}}\}_{f,i} =\begin{cases} 
       \frac{I \sigma_a(\nu_l)}{h \nu_l} & i>f~\textrm{(upper triangle)}\\
      \frac{I \sigma_{se}(\nu_l)}{h \nu_l} & i<f~\textrm{(lower triangle)}\\ 
   \end{cases}
\end{equation}
where $I$ is the intensity and the absorption and stimulated emission cross sections are~\cite{Demtroder2013}
\begin{widetext}
\begin{align}\label{eq:sigma_a}
\sigma_a(\nu_l) &=\sum_k  
\left(\frac{2J_k'+1}{2J_k''+1}\right)\frac{\lambda_k^2}{8\pi} A_{21}^v \times \frac{L(\Delta_k + \delta,\gamma)+L(\Delta_k - \delta,\gamma)}{2}\\
\sigma_{se}(\nu_l) &=\sum_k  
\frac{\lambda_k^2}{8\pi} A_{21}^v \times \frac{L(\Delta_k + \delta,\gamma)+L(\Delta_k - \delta,\gamma)}{2}
\end{align}
\end{widetext}

which ensures that the pumping of each transition individually satisfies detailed balance. The sum over $k$ runs over all dipole-allowed transitions (with selection rule $R_i=R_f$), and the sum of two detuned Lorentzians account for pumping from two counterpropagating beams (Fig.~\ref{fig:SI_pumping_model}). The line shape function,
\begin{equation}
    L(x,\gamma) = \frac{\gamma}{2 \pi}\frac{1}{x^2+(\gamma/2)^2},
\end{equation}
is an area-normalized Lorentzian. $\Delta_k = \nu_l - \nu_k$ is the detuning from the \textit{rest-frame} molecular resonance frequency of the $k$-th transition.  $\lambda_k=c/\nu_k$ is the wavelength of the $k$-th transition. Expressions for $\nu_k$ are given in Section~\ref{sec:transition_freqs}. No optical pumping of states in the vibrational reservoir is considered since those transitions will be far off resonance.

Finally, to conserve probability, the diagonals of $\mathbf{R_{OP}}$ are set to make the sum of columns zero. 

We have accounted for dephasing and lifetime broadening by manually incorporating an area-normalized, pressure-broadened homogeneous linewidth $\gamma$ for the frequency-dependent absorption cross section. Thus the direct effect of homogeneous broadening on the effective pumping rate is to depress the peak absorption cross section on resonance and spread it to neighbouring frequencies. This model ignores coherences in collision- or diffusion-induced population transfer, and assumes the radiation field only drives pairs of states at a time without inducing coherences between three or more levels.

To account for inhomogeneous broadening, we discretize the Doppler velocity profile $P(v_z) = g(\gamma=0,\sigma_D,\beta = 0,v_z/\lambda)$ (lineshape function $g$ is defined in Equation~\ref{eq:raut_fn}) into bins each spanning from $v_z$ to $v_z+dv_z(v_z)$. The $dv_z(v_z)$ are calculated to give constant partial integrated probability (made sufficiently small to give convergence in profile fitting)

\begin{equation}
    P(v_z)\textrm{d}v_z(v_z) = \begin{cases} 
      0.1 & \textrm{Q branch} \\
      0.03 & \textrm{R branch}
   \end{cases}
\end{equation}

For each bin we solve for the velocity-class resolved partial equilibrium population distributions $\mathbf{v_0}^{v_z}$ with $\delta=  v_z/\lambda$. Their weighted sum yields the total population distribution 
\begin{equation}
    \sum_{v_z} P(v_z)\textrm{d}v_z(v_z) \times \mathbf{v_0}^{v_z}=\mathbf{v_0}^{tot}
\end{equation} This sum assumes velocity classes are totally decoupled, a conclusion which is independently supported by the R-branch lineshape fitting (Section~\ref{sec:lineshape}). Finally, the full absorption spectrum is obtained by repeating this calculation for different values of the laser frequency $\nu_l$.

\section{Vibrational partition functions.}\label{ssec:vib partition function}
We calculate the vibrational partition functions for detailed balancing of the vibrational state transition matrix (Equation~\ref{eq:KV structure}). We assume a temperature of 150~K. The total partition function of C$_{60}$ vibrational modes \textit{i}, with degeneracy $g_i$ and vibrational energy spacing $\hbar \omega_i$, is 
\begin{align}
Z_{vib} &=\prod_i (Z_{1D}^i)^{g_i} \\
& =\prod_i  \frac{1}{(1- e^{-\hbar \omega_i/kT})^{g_i}}\\
&=2.74
\end{align}

The contributions to the partition function from the ground state and excited triply degenerate 1185~cm$^{-1}$ $T_{1u}(3)$ state are   
\begin{align}
Z_{v=0} &=1 \\
Z_{v=1} & =3\times e^{-1185~\textrm{cm}^{-1}/kT}\\
& =4\times 10^{-5}
\end{align}
The partition function of the reservoir is therefore approximately
\begin{align}
Z_{res} &\approx Z_{vib}-Z_{\nu=0}-Z_{\nu=1}\\
&=1.74
\end{align}

\section{Transition frequencies.}\label{sec:transition_freqs}
We present expressions for the C$_{60}$ transition frequencies. The spectroscopic constants $\Delta B=-1.8(1)\times10^{-7}~$cm$^{-1}$ and $\Delta D= -8.9(9)\times10^{-13}~$cm$^{-1}$ have been obtained from fitting to data in Ref.~\cite{Changala2019}, assuming the rotational temperature was fixed to 150~K. 

In the Q-branch:
\begin{equation}\label{eq: v=0 nu}
    \nu_k = 1185~\textrm{cm}^{-1} + \Delta B J_k'' (J_k''+1) - \Delta D (J_k'' (J_k''+1))^2
\end{equation}

In the R-branch:
\begin{equation}\label{eq: v=1 nu}
\nu_k = 1185~\textrm{cm}^{-1} +2 B (1-\zeta)(J_k''+1)
\end{equation}

\section{Data reduction and lineshape fitting for R branch.}\label{sec:lineshape}

To extract lineshape parameters and subtract the etalon background, we fit the raw R-branch data using a phenomenological lineshape function~\cite{Demtroder1995}:
\begin{widetext}

\begin{align}\label{eq:raut_fn}
    f(\gamma,\sigma_D,\beta,\Delta,s_0)&=C g(\gamma,\sigma_D,\beta,\Delta)/ \left(2\gamma' \right)\\
    g(\gamma,\sigma_D,\beta,\Delta)&= \textrm{Re}\left[\frac{\Phi(z)}{1-\sqrt{\pi}(\beta/\sqrt{2}\sigma_D) \Phi(z)}\right]\bigg/\left(\sqrt{2\pi}\sigma_D\right)
\end{align}
\end{widetext}

where
\begin{align}
    s&=s_0\frac{\gamma'^2}{\Delta^2+\gamma'^2}\\
    C&=\frac{\gamma'}{  B\left(1-\left(\frac{2\Delta}{A+B}\right)^2\right)^{1/2}}  \\
    A & = \left(\Delta^2+\gamma'^2\right)^{1/2}\\
    B&=\left(\Delta^2+\gamma'^2(1+2s)\right)^{1/2}\\
    \gamma'&= \gamma \frac{\sqrt{1+s_0}}{2}\\
    z&=\frac{\Delta + i \gamma/2 + \beta}{\sqrt{2} \sigma_{D}}\\
\end{align}

\begin{comment}
    \sigma_D&=\sigma_{D}'/\sqrt{8 ln(2)}\\
  \sigma_{D}' &= \sqrt{\frac{8 ln(2)  k_B T }{m c^2}}\nu_0
\end{comment}

$\Phi(z)$ is the Fadeeva function~\cite{Wang2011} and $\sigma_D=\sqrt{\frac{ k T }{m c^2}}\nu_0$ is the Doppler width. This was found to perfectly reproduce saturated absorption lineshapes from the rate equation model (Section~\ref{sec:rate_eqn_model}) in the intermediate regime $s_0\lesssim 1$, and $\gamma\sim \sigma_{D}\sqrt{8 \textrm{ln}(2)}$, where we operate. For C$_{60}$-Ar collisions, we found the best-fit $\beta$, the velocity-narrowing parameter, to be consistent with zero, confirming that different velocity classes can be treated independently. For all lighter gases, we expect $\beta=0$ to also hold at the same temperature because the momentum transferred per collision by a lighter BG must be less ($\langle \Delta p^2\rangle = \mu k T$).

We further constrained the relations $\gamma \equiv \frac{1}{\pi}\frac{P}{kT} \sigma_{PB} \overline{v}_{rel}$ and $s_0=i/i_{sat}$, where $i_{sat} = \kappa+\kappa_2 \times P^2$. Fitting isolated R-branch saturated lineshapes taken at pressures $P$ ranging from $100-400$~mTorr and intensities ranging from $2-25~$mW/mm$^2$ yields a set of $\sigma_{PB}$, $\kappa$, and $\kappa_2$ parameters unique to each BG species. These three parameters could then used to reproduce a family of background-free lineshapes for each BG species across arbitrary pressures and intensities for fitting to the rate equation model (Section~\ref{sec:rate_eqn_model}).

\section{Complete C$_{60}-$Ar fitted R-branch profiles.}\label{sec:C60-Ar_R_full}

\begin{figure}[htbp!]
    \centering
    \includegraphics[width=\columnwidth]{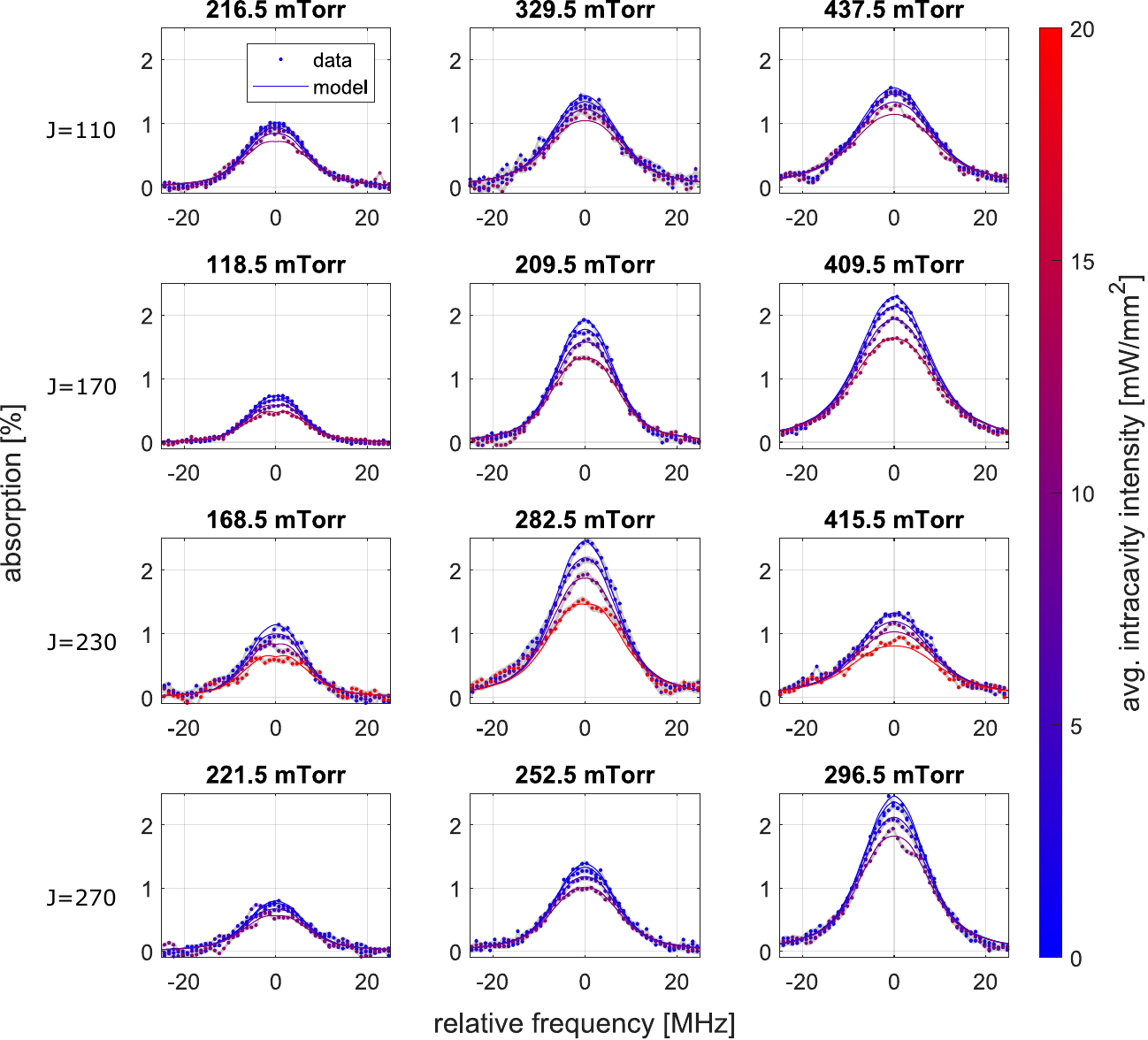}
    \caption{Full suite of $J$-dependent R-branch data and rate equation model fits for Ar-C$_{60}$ data}
    \label{fig:Ar_data}
\end{figure}

In addition to the C$_{60}-$Ar data shown in Fig.~\ref{fig:saturation_Q_R}C) and D), we measured a full suite of pressure, $J$-, and intensity-dependent R-branch absorption profiles at three pressures, at $J=110$, 170, 230, and 270. The data and corresponding fits derived from rate equation model (described in Section~\ref{sec:rate_eqn_model}) are shown in Fig.~\ref{fig:Ar_data}. 

\section{Complete C$_{60}-$BG fitted profiles.}\label{sec:fitted_all}

\begin{figure}[htbp!]
    \centering
    \includegraphics[width=\columnwidth]{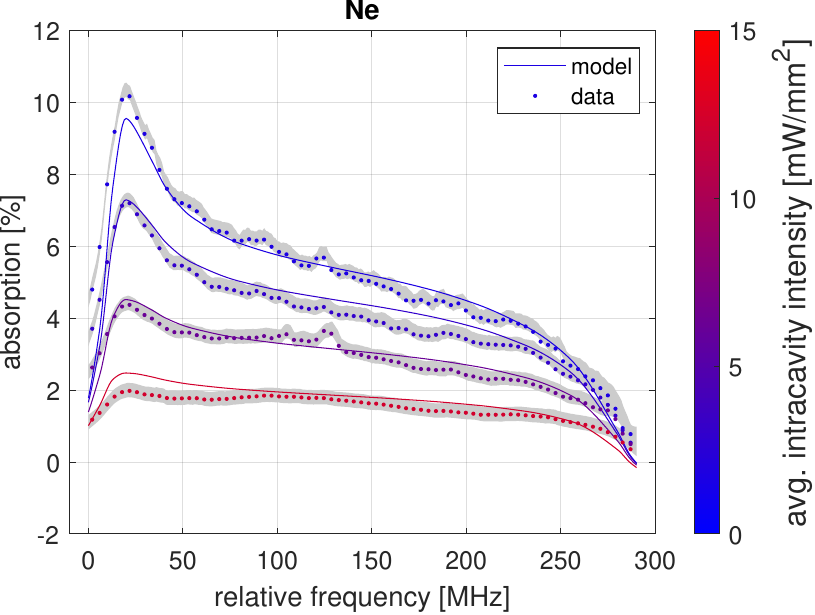}
    \caption{Best-fit Ne-C$_{60}$ Q branch profile}\label{fig:fitted_1}
\end{figure}

\begin{figure}[htbp!]
    \centering
    \includegraphics[width=\columnwidth]{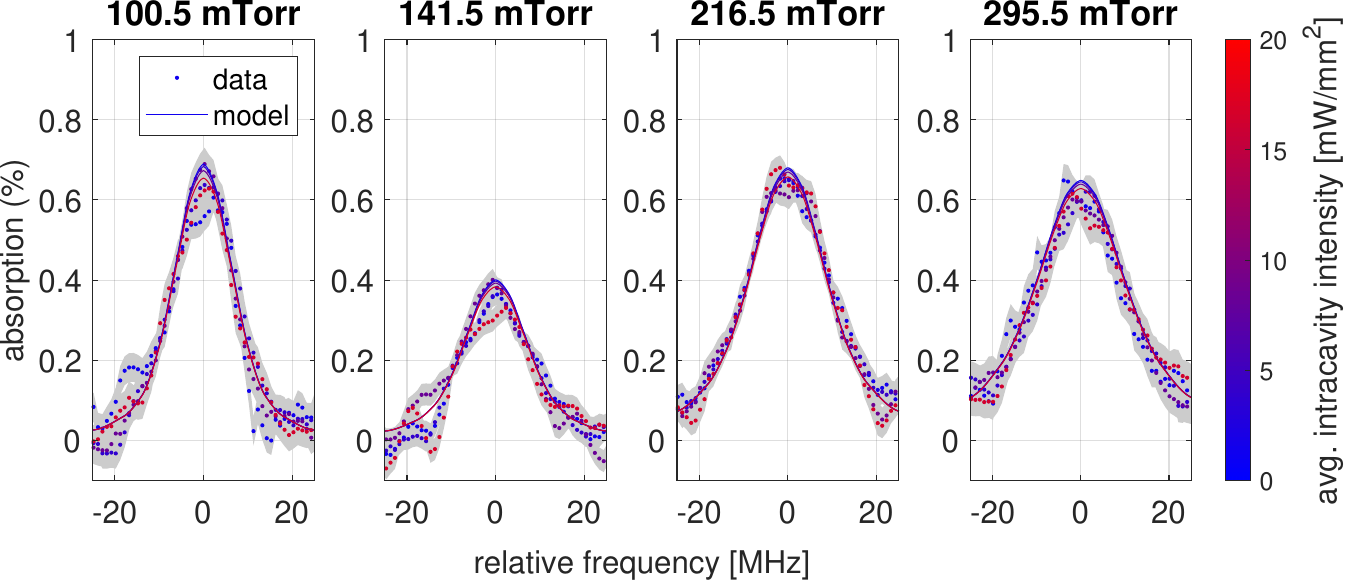}
    \caption{Best-fit H$_2$-C$_{60}$ R branch profiles}\label{fig:fitted_2}
\end{figure}

All fitted profiles for BG = Ne, He, D$_2$, and H$_2$ are shown in Figures~\ref{fig:fitted_1}-\ref{fig:fitted_2}.

\begin{figure}[htbp!]
    \centering
    \includegraphics[width=\columnwidth]{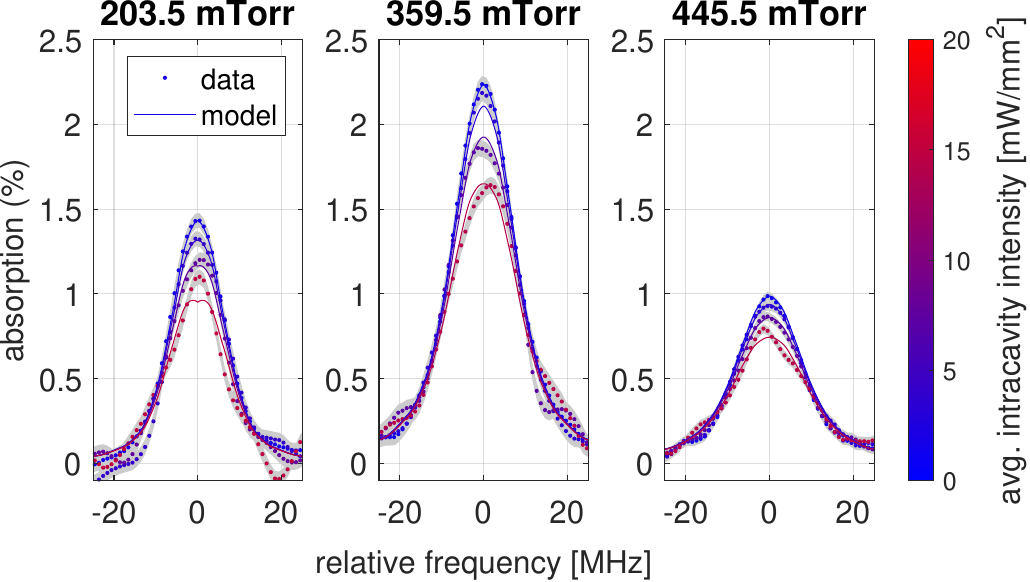}
    \caption{Best-fit Ne-C$_{60}$ R branch profiles}
\end{figure}

\begin{figure}[htbp!]
    \centering
    \includegraphics[width=\columnwidth]{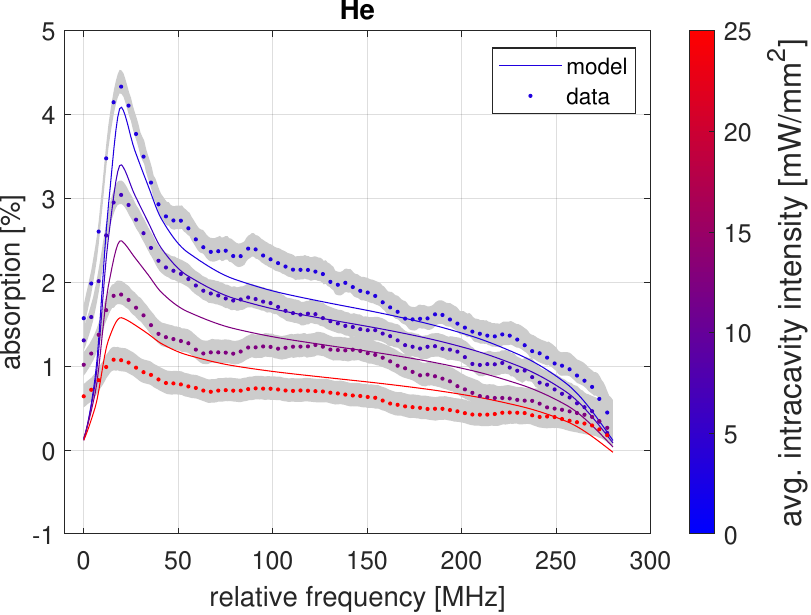}
    \caption{Best-fit He-C$_{60}$ Q branch profile}
\end{figure}

\begin{figure}[htbp!]
    \centering
    \includegraphics[width=\columnwidth]{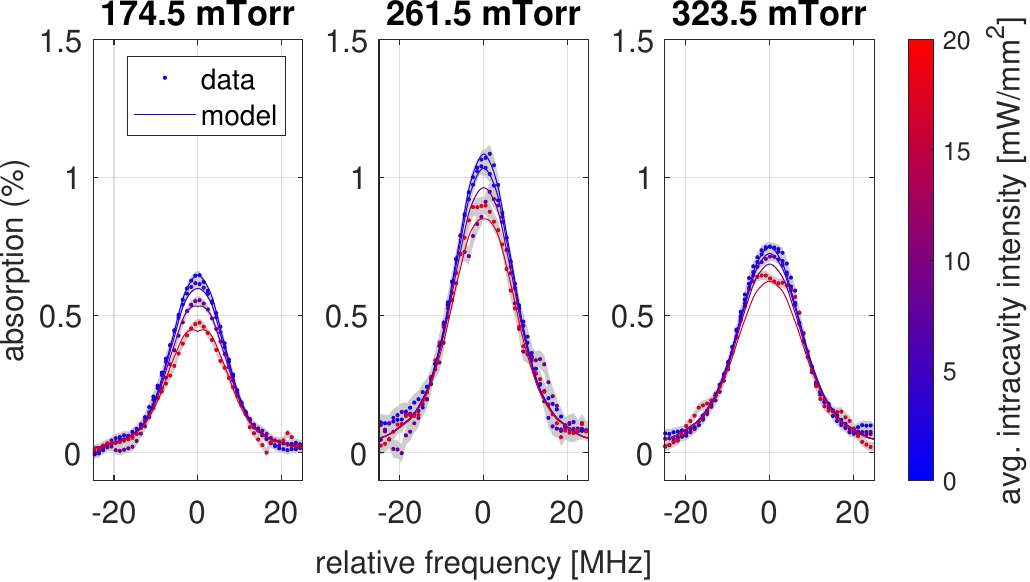}
    \caption{Best-fit He-C$_{60}$ R branch profiles}
\end{figure}

\begin{figure}[htbp!]
    \centering
    \includegraphics[width=\columnwidth]{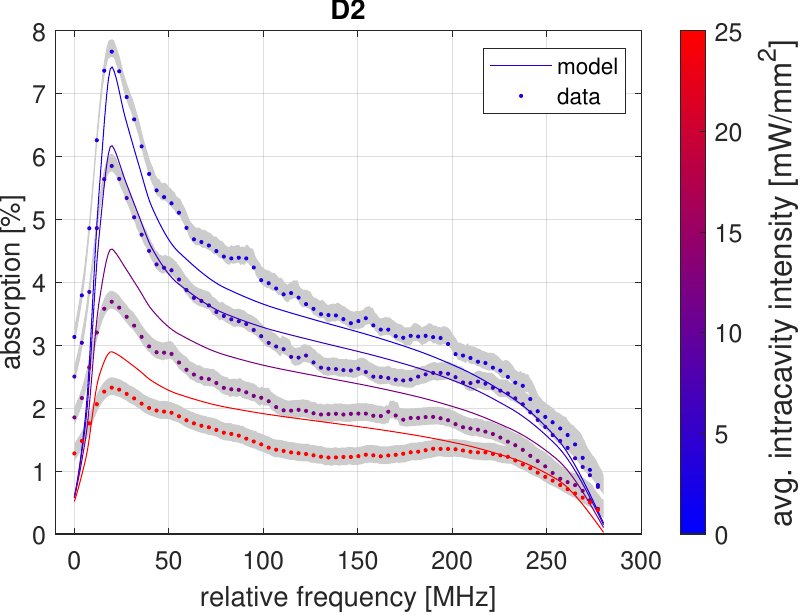}
    \caption{Best-fit D$_2$-C$_{60}$ Q branch profile}
\end{figure}

\begin{figure}[htbp!]
    \centering
    \includegraphics[width=\columnwidth]{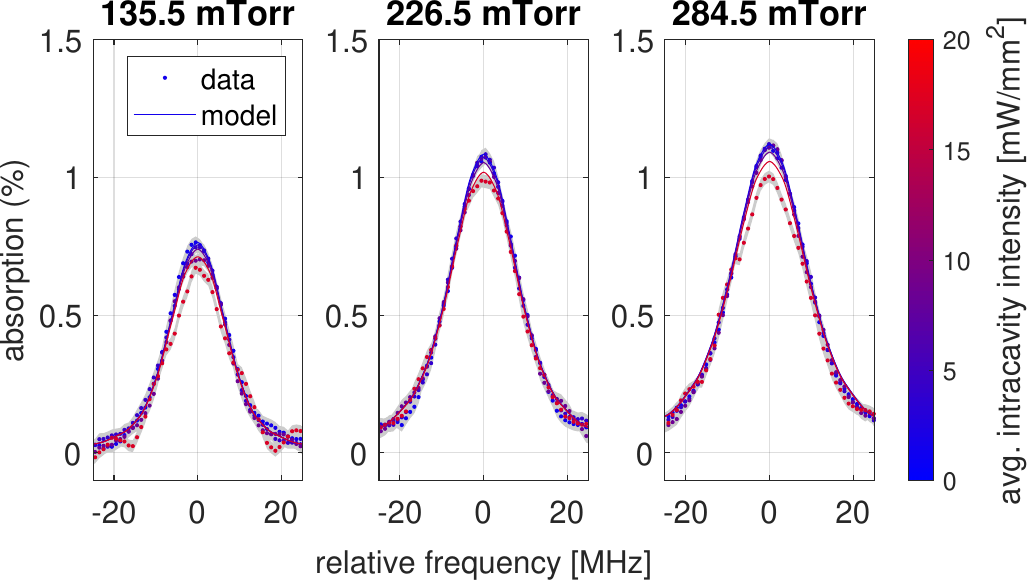}
    \caption{Best-fit D$_2$-C$_{60}$ R branch profiles}
\end{figure}

\begin{figure}[htbp!]
    \centering
    \includegraphics[width=\columnwidth]{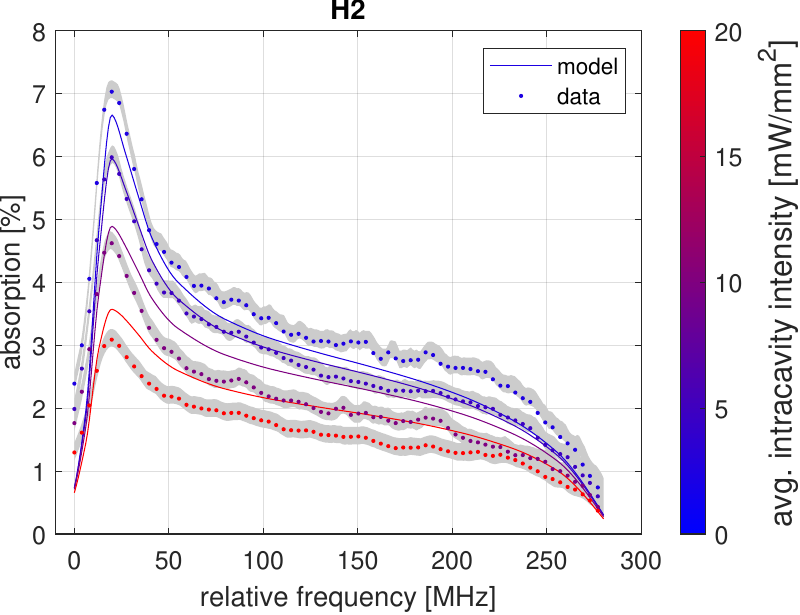}
    \caption{Best-fit H$_2$-C$_{60}$ Q branch profile}
\end{figure}

%Based on the low 20~mTorr vapor pressure~\cite{Abrefah1994}

\section{Calculating ground-state C$_{60}-$Ar and C$_{60}-$He potential energy surfaces.}\label{sec:calculate_PES}

The ground-state potential energy surface (PES) of the C$_{60}$-Ar and C$_{60}$-He  complexes are not available from the literature. We, therefore,  employed {\em ab-initio} density functional theory (DFT) to calculate  these potentials assuming a  rigid C$_{60}$ fullerene molecule with all C atoms at their equilibrium positions as functions of the position of  Ar or He. The counterpoise corrected DFT calculations are performed using the Gaussian 09 program~\cite{Gaussian09_RevE} employing the hybrid wB97XD functional~\cite{DFT_wb97xd} and the 6-31G(d,p) basis set. The  position of the rare-gas atom is defined in terms of   coordinates $(R,\theta,\phi)$, where $R$ is the separation of the noble gas atom  from the center of mass of C$_{60}$, and  polar $\theta$ and  azimuthal $\phi$ angles  are defined with respect to a Cartesian coordinate system with its $x$ and $z$ axes along a two-fold and five-fold symmetry axis of C$_{60}$, respectively.

The rigid $^{12}$C$_{60}$ molecule satisfies the  symmetries of the icosahedral group $I_h$. These symmetries restrict the allowed expansion coefficients when
the potentials $V(R,\theta,\phi)$ of C$_{60}$-Ar and C$_{60}$-He are expanded in spherical harmonic functions $C_{lm}(\theta,\phi)$. In fact,
we find to good approximation
\begin{equation}
%V(R,\theta,\phi)=\sum'_{l,m}V_{l,m}(R)C_{lm}(\theta,0)\cos (m\phi)\,,
V(R,\theta,\phi)=\sum_{l,m}^\prime \,V_{l,m}(R)\frac{C_{lm}(\theta,\phi)+C_{l-m}(\theta,\phi)}{2}
\,
\label{eq:expansion}
\end{equation}
where the prime on the  sum implies ${l=0}$, 6, 10, 12, 16, 18, 20, ${m=5 n}$,  ${n=0}$, 1, 2, 4, and $0\le m\le l$.
Finally, the fitted $V_{l,m}(R)$ are potential strengths.

\begin{figure} [htbp]
\centering
\includegraphics[scale=0.29,trim=0 0 0 0,clip]{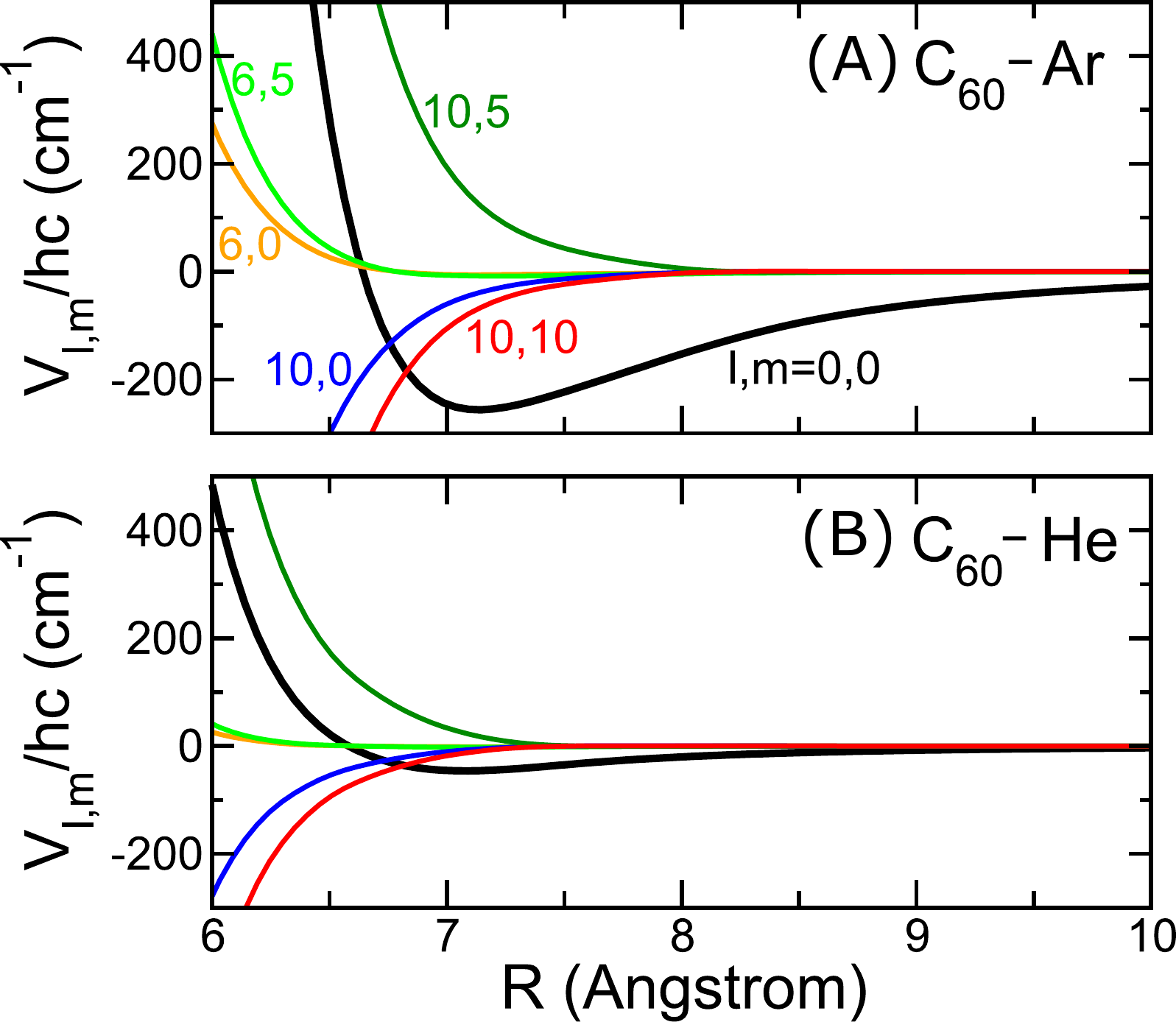}
\caption{Potential energy curves for  C$_{60}-$Ar and C$_{60}-$He. 
%Panels (a) and (b) show  contour plots  of the potentials {\color{blue}(in units of cm$^{-1}$)} as functions of angles $\theta$ and $\phi$ for  C$_{60}$-Ar 	and C$_{60}$-He, respectively. The separations between Ar and  C$_{60}$  and between He and  C$_{60}$ 	are their equilibrium bond lengths of $R_{\rm e}=7.2$~\AA\  and $R_{\rm e}=7.0$~\AA\ , respectively. 	Potential minima  are marked by white pentagons and hexagons corresponding to a rare-gas atom near a  five and six member ring of C$_{60}$, respectively.  Notice the very different energy of the two panels.
Panels (A) and (B) show strengths $V_{l,m}(R)$ for C$_{60}$-Ar and C$_{60}$-He as functions of $R$, respectively. The black curves correspond to the  isotropic potentials, which dominate in the atom-molecule bond.  The meaning of the line colors in panels (c) and (d) is the same.} 
\label{PESs}
\end{figure}

Panels (A) and (B) of Fig.~\ref{PESs} show the strongest strengths $V_{l,m}(R)$  as functions of $R$ for Ar and He, respectively. The isotropic strength $V_{0,0}(R)$ dominates over the other coefficients for $R>7$~\AA\ . Anisotropic terms have $l,m\ne 0,0$. The largest anisotropic contribution is that  for $l,m=10,5$ followed by those for  $10,10$ and $10,0$. In fact, these latter two  have the opposite sign from that for $l,m=10,5$. The remaining strengths shown in the graphs are even weaker. 
These anisotropic strengths determine the features of inelastic collisions of C$_{60}$ with noble gas atoms.

%\nocite{*}

\bibliography{refs,c60arhe_references,quenching_cs,SI}

\end{document}